\documentclass[sn-basic,iicol]{sn-jnl}

\usepackage{soul}
\jyear{2022}%

\begin{document}

\title[The blazar J1419$-$083]{Radio and mid-infrared properties of the blazar J1419$-$0838}

%%=============================================================%%
%% Prefix	-> \pfx{Dr}
%% GivenName	-> \fnm{Joergen W.}
%% Particle	-> \spfx{van der} -> surname prefix
%% FamilyName	-> \sur{Ploeg}
%% Suffix	-> \sfx{IV}
%% NatureName	-> \tanm{Poet Laureate} -> Title after name
%% Degrees	-> \dgr{MSc, PhD}
%% \author*[1,2]{\pfx{Dr} \fnm{Joergen W.} \spfx{van der} \sur{Ploeg} \sfx{IV} \tanm{Poet Laureate} 
%%                 \dgr{MSc, PhD}}\email{iauthor@gmail.com}
%%=============================================================%%

\author*[1,2]{\fnm{Krisztina} \sur{Perger}}\email{perger.krisztina@csfk.org}

\author[1,2,3]{\fnm{S\'{a}ndor} \sur{Frey}}\email{frey.sandor@csfk.org} 

\author[4,5,1,2]{\fnm{Krisztina \'{E}va} \sur{Gab\'{a}nyi}}\email{k.gabanyi@astro.elte.hu} 

\affil*[1]{\orgdiv{Konkoly Observatory}, \orgname{ELKH Research Centre for Astronomy and Earth Sciences}, \orgaddress{\street{Konkoly Thege Mikl\'{o}s \'{u}t 15-17}, \city{Budapest}, \postcode{H-1121},  \country{Hungary}}}

\affil[2]{\orgdiv{CSFK}, \orgname{MTA Centre of Excellence},  \orgaddress{\street{Konkoly Thege Mikl\'{o}s \'{u}t 15-17}, \city{Budapest}, \postcode{H-1121},  \country{Hungary}}}

\affil[3]{\orgdiv{Institute of Physics and Astronomy}, \orgname{ELTE E\"{o}tv\"{o}s Lor\'{a}nd University}, \orgaddress{\street{P\'{a}zm\'{a}ny P\'{e}ter s\'{e}t\'{a}ny 1/A}, \city{Budapest}, \postcode{H-1117},  \country{Hungary}}}

\affil[4]{\orgdiv{Department of Astronomy, Institute of Physics and Astronomy}, \orgname{ELTE E\"{o}tv\"{o}s Lor\'{a}nd University}, \orgaddress{P\'{a}zm\'{a}ny P\'{e}ter s\'{e}t\'{a}ny 1/A}, \city{Budapest}, \postcode{H-1117}, \country{Hungary}}

\affil[5]{\orgname{ELKH--ELTE Extragalactic Astrophysics Research Group}, \orgaddress{P\'{a}zm\'{a}ny P\'{e}ter s\'{e}t\'{a}ny 1/A}, \city{Budapest}, \postcode{H-1117}, \country{Hungary}}

\abstract{The radio quasar NVSS~J141922$-$083830 (J1419$-$0838) was initially classified as an uncategorised blazar-type object, following its detection in the $\gamma$-ray band with the \textit{Fermi} space telescope. Later, using multi-waveband observations and modeling, its was found to be a flat-spectrum radio quasar (FSRQ). However, its radio emission has never been discussed in depth in the literature. Here we present a detailed analysis on the radio properties of J1419$-$0838 using archival interferometric imaging data at pc and kpc scales. We conclude that the flux density variations, the flat radio spectrum, the compact nature of the quasar structure at all scales, and the relativistic Doppler enhancement of the radio emission all support the previous classification as an FSRQ. We also investigated the short- and long-term mid-infrared (MIR) light curve of the quasar based on observations by the \textit{Wide-field Infrared Survey Explorer} satellite, and found that there is significant variability on time-scales of days as well as years. Comparison of the MIR light curve to the times of previously reported $\gamma$-ray and optical flares shows no clear correlation between the events at different wavebands.}


%%\pacs[JEL Classification]{D8, H51}

%%\pacs[MSC Classification]{35A01, 65L10, 65L12, 65L20, 65L70}

\maketitle

\section{Introduction}\label{sec1}
Active galactic nuclei (AGNs) are compact regions in the centres of galaxies, producing luminous emission in the entire electromagnetic spectrum. The most powerful AGNs are believed to have their emission relativistically enhanced because of their jet orientation close to the line of sight to the observer, and are called blazars. There are two main sub-classes of blazars, BL Lac objects and flat-spectrum radio quasars (FSRQs). The radio-loud quasar NVSS~J141922$-$083830 (J1419$-$0838 hereafter; right ascension $14^\mathrm{h} 19^\mathrm{m} 22.556072^\mathrm{s}$, declination $-08^{\circ} 38^{\prime} 32.14018^{\prime\prime}$ taken from the Radio Fundamental Catalogue version 2022c\footnote{\url{http://astrogeo.org/sol/rfc/rfc_2022c/}})  has a reported redshift of $z=0.90$ \citep{2022MNRAS.517.5791B,2022MNRAS.516.5702O}. We note that the VLBI coordinates coincide with the optical position measured by Gaia within $0.3$~mas \citep{2016A&A...595A...1G,2022arXiv220800211G}, i.e., within the formal uncertainties. The quasar was labeled as a blazar of uncertain type in the \textit{Fermi} Large Area Telescope (LAT) Third AGN Catalog \citep[3LAC,][]{2015ApJ...810...14A}, with no other distinction on its nature. Its X-ray, $\gamma$-ray, optical, and near-infrared properties were further investigated by \citet{2022MNRAS.517.5791B} and \citet{2022MNRAS.516.5702O}, including spectral energy distribution (SED) modeling. These studies led to the classification of J1419$-$0838 as an FSRQ, without actually investigating whether its radio spectrum is flat or not. This new classification is also reflected in the \textit{Fermi}-LAT Fourth AGN Catalog \citep[4LAC,][]{2022arXiv220912070T}. However, these works do not take the radio properties into account in the classification process and the analysis of the source. Here we provide a detailed account on what is known about J1419$-$0838 in the radio, using archival single-dish and interferometric data with low and high angular resolutions, covering large and small spatial scales, respectively. We also analyse mid-infrared (MIR) flux density monitoring observations from the \textit{Wide-field Infrared Survey Explorer (WISE)} satellite.

Adopting a standard $\Lambda$CDM cosmological model with $H_0=70$~km~s$^{-1}$~Mpc$^{-1}$, $\Omega_\mathrm{M}=0.3$, and $\Omega_\Lambda = 0.7$, 1 milliarcsecond (mas) angular size corresponds to 7.79~pc projected linear size at the redshift of J1419$-$0838. The quasar is at the luminosity distance $D_\mathrm{L}=5800.9$~Mpc \citep{2006PASP..118.1711W}.

\section{Observations and data reduction}

\subsection{Archival radio data}

We collected non-simultaneous radio flux density measurements for J1419$-$0838 from the literature and various public catalogues.  
The details and references are listed in Table~\ref{tab:spectrum}. The observing frequencies span a wide range of more than 3 orders of magnitude, from 76~MHz to 285~GHz.

\subsection{VLBI imaging and model-fitting}

To determine the physical parameters of the jet in the radio quasar J1419$-$0838, we turned to the highest-resolution data and analysed available archival very long baseline interferometry (VLBI) measurements. The data were obtained with the US National Radio Astronomy Observatory (NRAO) Very Long Baseline Array (VLBA) at two different frequencies, $4.3$ and $7.6$~GHz, in the framework of the 7th and 8th runs of the VLBA Calibrator Survey \citep[PI: L. Petrov, project codes: BP171a3, BP171a5, BP177b, and BP177c,][]{2021AJ....161...14P} at four different epochs in 2013 and 2014. The details of the observations are given in Table~\ref{tab:vlbaobs}. The calibrated interferometric visibility data were downloaded from the Astrogeo database\footnote{\url{http://astrogeo.org/cgi-bin/imdb_get_source.csh?source=J1419-0838}}. The short snapshot observations were made in right circular polarisation, with $32$~MHz bandwidth in each of the 8 intermediate frequency channels (IFs), leading to a total bandwidth of $256$~MHz. The calibrated data sets were produced by the \textsc{pima} software \citep[v1.28 and v2.08,][]{2011AJ....142...35P}. We performed hybrid mapping in \textsc{difmap} \citep{difmap}, using subsequent cycles of \textsc{clean} deconvolution \citep{clean} and phase and amplitude self-calibration. Finally, we fitted circular Gaussian brightness distribution model components \citep{modelfit} directly to the self-calibrated visibility data, to describe the brightness distribution of J1419$-$0838 quantitatively.

\subsection{\textit{WISE} infrared data}

To address the infrared flux density variability, we collected data from \textit{WISE} observations performed in the \textit{WISE} and \textit{NEOWISE} missions \citep{2010AJ....140.1868W,2014ApJ...792...30M}. 

We downloaded data from the NASA/IPAC Infrared Science Archive\footnote{\url{https://irsa.ipac.caltech.edu/frontpage/}} obtained by the {\it WISE} satellite at the position of J1419$-$0838 within a search radius of $6''$. This resulted in $32$ and $205$ points within the framework of the original {\it WISE} mission (between 2009 December and 2011 February) and the \textit{NEOWISE} Reactivation Mission (from 2013 December up until the beginning of 2021), respectively. These include measurements in all four bands of the \textit{WISE} satellite, $3.4\,\mu$m ($W1$), $4.6\,\mu$m ($W2$), $12\,\mu$m ($W3$), and $22\,\mu$m ($W4$) at the beginning of the mission, and observations taken only at the two shorter wavelengths after the depletion of the cooling material. Following the Explanatory Supplement to the \textit{WISE} All-Sky Data Release Products\footnote{\url{https://wise2.ipac.caltech.edu/docs/release/allsky/expsup/index.html}}, we flagged those measurements which have lower photometric quality (given in column {\verb!ph_qual!}), may be affected by scattered moonlight (denoted in column {\verb!moon_masked!}), were obtained when the satellite's separation from the South-Atlantic Anomaly was $\leq 5^\circ$ (denoted in column {\verb!saa_sep!}), or were reported to be extended. Finally, we retained $195$ points, more than $80$\,\% of the measurements. We converted the magnitude measurements to flux densities using the appropriate conversion factors in the given band (for a power-law spectrum with an exponent of $-1$) listed in the Explanatory Supplement to the WISE All-Sky Data Release Products.

\section{Results and discussion}

\subsection{Radio spectrum}

Figure~\ref{fig:spectrum} displays the radio spectrum of J1419$-$0838 compiled from flux density measurements listed in Table~\ref{tab:spectrum}. To characterise the spectrum at both large (kpc) and small (pc) scales, we fitted power-law spectra\footnote{As there are significant differences in flux density values in the GLEAM catalogue \citep{2017MNRAS.464.1146H} between adjacent frequency bands observed at the same epoch, we excluded the MWA data points from the spectrum fit, although they are shown in Table~\ref{tab:spectrum} and in Fig.~\ref{fig:spectrum}. We note that the spectral index values determined with and without the MWA data points are in agreement.} in the form $S_\nu \propto \nu^\alpha$, where $S$ is the flux density, $\nu$ the observing frequency, and $\alpha$ the spectral index. The best-fit parameters and spectral slopes for the total $(S)$ and pc-scale compact flux densities $(S_\mathrm{VLBI})$ are shown in Fig.~\ref{fig:spectrum}.

With the inclusion of all data points, the quasar shows a flat radio-to-submm spectrum with a spectral index of $\alpha=0.15\pm0.36$. The flux densities obtained from the highest-resolution VLBI observations suggest an inverted spectrum, with somewhat steeper slope ($\alpha_\mathrm{VLBI}=0.48\pm0.13$).

\subsection{Radio flux density variability}

Most of the measurements shown in Fig.~\ref{fig:spectrum} are non-simultaneous, and flux density variability with time causes scatter in the spectral plot. Considering flux densities from the high-resolution VLBA observations, the quasar core shows some variability in the flux densities, up to $16\%$ at both frequency bands, on even a monthly timescale (Table~\ref{tab:vlbamodel}). There is further evidence when comparing the 2.3-GHz flux densities obtained from observations by the Parkes telescope prior to 2010 and the VLBA in 2018, as the latter shows a $14\%$ increase, despite the better angular resolution where any large-scale extended emission should be resolved out and thus remain undetected. The variability is also apparent from sub-mm observations by the Atacama Large Millimeter/submillimeter Array (ALMA) at $\sim95$, $233$, and $235$~GHz frequencies, performed at different epochs. These effects are well presented in the radio-to-submm spectrum (Fig.~\ref{fig:spectrum}) and the VLBI light curve (Fig.~\ref{fig:vlba_lc}).

\subsection{Large-scale radio structure}

In Fig.~\ref{fig:largescale}, we show the large-scale structure of the quasar from observations with the Murchison Widefield Array (MWA) and the Karl G. Jansky Very Large Array (VLA), obtained in the framework of the Galactic and Extra-Galactic All-SKY MWA Survey \citep[GLEAM,][]{2015PASA...32...25W,2017MNRAS.464.1146H}, the NRAO VLA Sky Survey \citep[NVSS,][]{1998AJ....115.1693C}, and the VLA Sky Survey \citep[VLASS,][]{2021ApJS..255...30G}. The NVSS radio image was obtained using the postage stamp server of NRAO\footnote{\url{https://www.cv.nrao.edu/nvss/postage.shtml}}, while the VLASS and GLEAM image data were obtained from the Canadian Initiative for Radio Astronomy Data Analysis\footnote{\url{http://cutouts.cirada.ca/}} (CIRADA) cutout service. The position of the quasar is denoted by its VLBI coordinates obtained from Radio Fundamental Catalogue. Contours of the radio maps from observations in the GLEAM and NVSS surveys superposed on the VLASS first-epoch radio image are shown in the left panel. For better visibility, we show the central $30^{\prime\prime} \times 30^{\prime\prime}$ area of the higher-resolution VLASS radio map in the right panel, along with the accurate VLBI position. The source shows an unresolved, featureless single-component structure at all resolutions and frequencies at kpc scales.

\subsection{Inner structure}

The naturally weighted \textsc{clean} maps of VLBA data sets resulting from the imaging are shown in Fig.~\ref{fig:maps}. Parameters of the maps and the fitted model components, along with the derived physical parameters are listed in Table~\ref{tab:vlbamodel}. We calculated errors for the model parameters following \cite{2008AJ....136..159L}, with an additional $5\%$ absolute amplitude calibration uncertainty in the flux densities.

We found that the quasar appears as a single-component compact `core' (i.e., the synchrotron self-absorbed base of the jet) at all VLBI epochs, which is further confirmed by the model-fitting results.

For all four available epochs with VLBI observations, the visibility data can be best fitted with a single Gaussian brightness distribution model component (Fig.~\ref{fig:maps} and Table~\ref{tab:vlbamodel}). We calculated the brightness temperature values using the following equation \citep[e.g.][]{1982ApJ...252..102C}:
\begin{equation}
  T_\mathrm{b} = 1.22\times10^{12}(1+z)\frac{S}{\vartheta^2\nu^2}~\mathrm{K},
\end{equation}
where $S$ is the flux density of the core model component measured in Jy, $\vartheta$ the component diameter (full width at half-maximum, FWHM) in mas, and $\nu$ the observing frequency expressed in GHz.

Using these measured brightness temperature values that are also given in Table~\ref{tab:vlbamodel}, we determined Doppler factors for each epoch as
\begin{equation}
  \delta=\frac{T_\mathrm{b}}{T_\mathrm{b,int}},
\end{equation}
where the $T_\mathrm{b,int} \approx 5\times10^{10}$~K intrinsic brightness temperature corresponds to the energy state where the radiating particles and the magnetic field are in equipartition \citep{1994ApJ...426...51R}. We found that the measured brightness temperatures all exceed the theoretical intrinsic equipartiton value (i.e., $\delta > 1$), thus the radio emission of the quasar core shows relativistic Doppler enhancement at the epochs of all VLBI observations (Table~\ref{tab:vlbamodel}), with $\delta$ values ranging between $2.6$ and $8.6$.

Monochromatic radio power values characteristic to the mas-scale radio `core' emission were also calculated using the following expression:
\begin{equation}
  P=4\pi S D_\mathrm{L}^2(1+z)^{-\alpha-1},
\end{equation}
where $\alpha=\alpha_\mathrm{VLBI}=0.48$ is the spectral index determined from the VLBA observations. We found that the monochromatic powers are $P_\mathrm{4.3GHz}=(2.9\pm0.5)\cdot10^{26}$~W~Hz$^{-1}$ and $P_\mathrm{7.6GHz}=(3.5\pm0.6)\cdot10^{26}$~W~Hz$^{-1}$ for the 4.3- and 7.6-GHz VLBA observations, respectively. Here we give the mean value of the four observations made at different epochs, stating the uncertainties as $3\sigma$ standard deviation of the powers at each frequency band.

To assess the compactness of the radio emission region, we can compare the total (arcsec-scale) flux densities with those originating from the mas-scale structure only. Given the time variability and the uncertainties of the individual measurements, the total flux densities (Table~\ref{tab:spectrum} and Fig.~\ref{fig:spectrum}) interpolated to the VLBA observing frequencies (4.3 and 7.6-GHz) are not significantly higher than the values obtained for the mas-scale structure (Table~\ref{tab:vlbamodel}). Therefore, we can conclude that the radio emission of J1419$-$0838 is dominated by the inner pc-scale structure. In other words, the VLBI observations do not resolve out much of the emission measured by lower-resolution instruments.

\subsection{Mid-infrared variability}

Prior to the detailed $\gamma$-ray--optical analysis of J1419$-$0838 \citep{2022MNRAS.517.5791B}, a large optical flare \citep{2015ATel.7133....1L} and a 5-magnitude  brightening was reported at {$1.63~\mu\mathrm{m}$ for this quasar in 2015 February  with the Cananea Near-Infrared Camera of the telescope at the Guillermo Haro Observatory \citep{2015ATel.7168....1C}.

To compare the MIR flux density changes with the $\gamma$-ray and optical flares reported by \citet{2022MNRAS.517.5791B}, we show the 100--800~MeV \textit{Fermi}-LAT light curve \citep[LCR,][]{2017ApJ...846...34A,2021ATel15110....1F} with grey solid and black dotted stripes indicating the aforementioned flaring events in Fig.~\ref{fig:wise_lc} (upper panel). 
Although the cadence of the \textit{WISE} observations is around 3--4 months, a clear long-term variability can be detected in the MIR light curve of the quasar at 3.4 and 4.6~$\mu$m ($W1$ and $W2$ filters) on a monthly time scale (middle panel in Fig.~\ref{fig:wise_lc}). Moreover, in several \textit{WISE} epochs, there is a clear intra-day variability pattern during the observations. Two `peaking' events were also identified in 2015 January and 2017 January, with $\sim30\%$ brightening and fading of the flux densities in $W1$ and $W2$. These particular mission phases were chosen for display because they are the ones that show the clearest intra-day variations, also including a peak during the brightness changes. Moreover, as we discuss later, these data highlight that the MIR light curve peaks have no apparent correlation to the $\gamma$-ray and optical flaring events. Apart from the two intra-day peaking events in 2015 and 2017 January, an overall brightening with respect to the quiescent flux density level is apparent at these epochs. Although the third $\gamma$-ray--optical flare \citep{2022MNRAS.517.5791B} occurred within two months of the first MIR brightening episode in 2015, there are no other indications of any correlation/connection between the changes in the $\gamma$-ray--optical and the MIR light curves.

In a sample of more than $10\,000$ AGNs, \citet{2022ApJ...927..107S} found a strong correlation between the variability of the MIR magnitudes and the $W1-W2$ colour, concluding that the variations are unlikely to originate from the contribution of the host galaxy, but likely to arise from the dusty torus of the AGNs. J1419$-$0839 shows a similar trend as the Pearson's correlation coefficient calculated from the available \textit{WISE} data between the $W1$ magnitude and $W1-W2$ colour is $0.99$, with $p<0.001$. Significant brightness changes within a day, however, cannot be explained with the contribution of the surrounding torus.

The intra-day and monthly magnitude variability, combined with the changes of the optical spectral lines \citep{2022MNRAS.517.5791B} imply similarities between J1419$-$0839 and the so-called changing-look AGNs, for which the variations are believed to be induced by the rise in the accretion rate \citep{2017ApJ...846L...7S}. Another possibility is that the variability in the MIR emission is caused by the relativistically boosted radio jet \citep[e.g.][]{2012ApJ...759L..31J, 2018MNRAS.477.5127Y,2018rnls.confE..42G,2018RNAAS...2..130G} which the quasar J1419$-$0839 does have.

\section{Summary and conclusions}

NVSS~J141922$-$083830 (J1419$-$0838) was previously classified as a flat-spectrum radio quasar, based on its $\gamma$-ray, X-ray, optical, and near-infrared properties. However, no detailed analysis was published about its radio emission. Here we utilised publicly available archival radio data at pc and kpc scales to characterise its radio emission, both quantitatively and qualitatively. The quasar has a single compact mas-scale `core' component at all five VLBI epochs between 2013 and 2018, and shows high measured brightness temperature values that indicate Doppler-enhanced emission from a relativistic jet pointing close to the line of sight. The radio source has indeed a flat (inverted) spectrum. Based on the radio data, we confirm that the quasar belongs to the blazar subclass of FSRQs.

Based on \textit{WISE} data, we found that the MIR light curve of the quasar shows both short- and long-term variability, most likely connected with AGN phenomena. Neither the intra-day, nor the monthly light variations indicate obvious correlation with the flaring events identified in the optical and $\gamma$-ray light curves.

\backmatter

\bmhead{Acknowledgments}
This research has made use of the CIRADA cutout service at \url{http://cutouts.cirada.ca}, operated by the Canadian Initiative for Radio Astronomy Data Analysis (CIRADA). CIRADA is funded by a grant from the Canada Foundation for Innovation 2017 Innovation Fund (Project 35999), as well as by the Provinces of Ontario, British Columbia, Alberta, Manitoba and Quebec, in collaboration with the National Research Council of Canada, the US National Radio Astronomy Observatory and Australia’s Commonwealth Scientific and Industrial Research Organisation.

VLBA is operated by the National Radio Astronomy Observatory, which is a facility of the National Science Foundation, and operated under cooperative agreement by Associated Universities, Inc. The use of the VLBA under the US Naval Observatory's time allocation is acknowledged. This work supports USNO's ongoing research into the celestial reference frame and geodesy. These data were retrieved from the VLBA public archive. 

This research has made use of the VizieR catalogue access tool, CDS, Strasbourg, France (DOI: 10.26093/cds/vizier). The original description  of the VizieR service was published in \citet{2000A&AS..143...23O}.
%2000, A\&AS 143, 23
 
This scientific work makes use of the Murchison Radio-astronomy Observatory, operated by CSIRO. We acknowledge the Wajarri Yamatji people as the traditional owners of the Observatory site. Support for the operation of the MWA is provided by the Australian Government (NCRIS), under a contract to Curtin University administered by Astronomy Australia Limited. We acknowledge the Pawsey Supercomputing Centre which is supported by the Western Australian and Australian Governments.

This research has made use of material from the Bordeaux VLBI Image Database (BVID). This database can be reached at \url{http://bvid.astrophy.u-bordeaux.fr/}.
% S and X band data from VLBA observations in 2018

This publication made use of the Astrogeo VLBI FITS image database (\url{http://astrogeo.org/vlbi_images/}) and we thank Leonid Petrov for placing observations online prior to publication.

This publication makes use of data products from the Wide-field Infrared Survey Explorer, which is a joint project of the University of California, Los Angeles, and the Jet Propulsion Laboratory/California Institute of Technology, funded by the National Aeronautics and Space Administration.

\section*{Statements and declarations}

\subsection*{Funding}
This research was supported by the Hungarian National Research, Development and Innovation Office (NKFIH), grant number OTKA K134213.

\subsection*{Declaration of Competing Interest}
The authors have no relevant financial or non-financial interests to disclose.

\subsection*{Author contribution}
The study conception and methodology was proposed by S\'andor Frey.
Collection of archival radio data, analysis, data visualisation were done, and the first draft of the manuscript was written by Krisztina Perger. Collection and initial flagging of infrared data were performed by Krisztina \'E. Gab\'anyi. All authors contributed to the previous versions of the draft, read and approved the content before the submission of the final manuscript.

\subsection*{Data availability}
Calibrated VLBI visibility data sets utilised in this work can be obtained from the Astrogeo Image Database (\url{http://astrogeo.org/}). Data from the spectral analysis are available from the VizieR catalogue access tool (\url{https://vizier.cds.unistra.fr/}). \textit{Fermi}-LAT data are available in the \textit{Fermi} LAT Light Curve Repository (\url{https://fermi.gsfc.nasa.gov/ssc/data/access/lat/LightCurveRepository/}).
Large-scale radio maps can be downloaded from the CIRADA Image Cutout Web Service (\url{http://cutouts.cirada.ca/}).

\bibliography{references.bib}% common bib file

\begin{thebibliography}{39}
\providecommand{\natexlab}[1]{#1}
\providecommand{\url}[1]{{#1}}
\providecommand{\urlprefix}{URL }
\providecommand{\doi}[1]{\url{https://doi.org/#1}}
\providecommand{\eprint}[2][]{\url{#2}}
 \bibcommenthead

\bibitem[{{Abdollahi} et~al(2017){Abdollahi}, {Ackermann}, {Ajello}, {Albert},
  {Baldini}, {Ballet}, {Barbiellini}, {Bastieri}, {Becerra Gonzalez},
  {Bellazzini}, {Bissaldi}, {Blandford}, {Bloom}, {Bonino}, {Bottacini},
  {Bregeon}, {Bruel}, {Buehler}, {Buson}, {Cameron}, {Caragiulo}, {Caraveo},
  {Cavazzuti}, {Cecchi}, {Chekhtman}, {Cheung}, {Chiaro}, {Ciprini}, {Conrad},
  {Costantin}, {Costanza}, {Cutini}, {D'Ammando}, {de Palma}, {Desai},
  {Desiante}, {Digel}, {Di Lalla}, {Di Mauro}, {Di Venere}, {Donaggio},
  {Drell}, {Favuzzi}, {Fegan}, {Ferrara}, {Focke}, {Franckowiak}, {Fukazawa},
  {Funk}, {Fusco}, {Gargano}, {Gasparrini}, {Giglietto}, {Giomi}, {Giordano},
  {Giroletti}, {Glanzman}, {Green}, {Grenier}, {Grove}, {Guillemot}, {Guiriec},
  {Hays}, {Horan}, {Jogler}, {J{\'o}hannesson}, {Johnson}, {Kocevski}, {Kuss},
  {La Mura}, {Larsson}, {Latronico}, {Li}, {Longo}, {Loparco}, {Lovellette},
  {Lubrano}, {Magill}, {Maldera}, {Manfreda}, {Mayer}, {Mazziotta},
  {Michelson}, {Mitthumsiri}, {Mizuno}, {Monzani}, {Morselli}, {Moskalenko},
  {Negro}, {Nuss}, {Ohsugi}, {Omodei}, {Orienti}, {Orlando}, {Paliya},
  {Paneque}, {Perkins}, {Persic}, {Pesce-Rollins}, {Petrosian}, {Piron},
  {Porter}, {Principe}, {Rain{\`o}}, {Rando}, {Razzano}, {Razzaque}, {Reimer},
  {Reimer}, {Sgr{\`o}}, {Simone}, {Siskind}, {Spada}, {Spandre}, {Spinelli},
  {Stawarz}, {Suson}, {Takahashi}, {Tanaka}, {Thayer}, {Thompson}, {Torres},
  {Torresi}, {Tosti}, {Troja}, {Vianello}, and {Wood}}]{2017ApJ...846...34A}
{Abdollahi} S, {Ackermann} M, {Ajello} M, et~al (2017) {The Second Catalog of
  Flaring Gamma-Ray Sources from the Fermi All-sky Variability Analysis}. \apj
  846(1):34. \doi{10.3847/1538-4357/aa8092},
  {\href{https://arxiv.org/abs/1612.03165}{{https://arxiv.org/abs/arXiv:1612.03165}}}
  {[astro-ph.HE]}

\bibitem[{{Ackermann} et~al(2015){Ackermann}, {Ajello}, {Atwood}, {Baldini},
  {Ballet}, {Barbiellini}, {Bastieri}, {Becerra Gonzalez}, {Bellazzini},
  {Bissaldi}, {Blandford}, {Bloom}, {Bonino}, {Bottacini}, {Brandt}, {Bregeon},
  {Britto}, {Bruel}, {Buehler}, {Buson}, {Caliandro}, {Cameron}, {Caragiulo},
  {Caraveo}, {Carpenter}, {Casandjian}, {Cavazzuti}, {Cecchi}, {Charles},
  {Chekhtman}, {Cheung}, {Chiang}, {Chiaro}, {Ciprini}, {Claus},
  {Cohen-Tanugi}, {Cominsky}, {Conrad}, {Cutini}, {D'Abrusco}, {D'Ammando}, {de
  Angelis}, {Desiante}, {Digel}, {Di Venere}, {Drell}, {Favuzzi}, {Fegan},
  {Ferrara}, {Finke}, {Focke}, {Franckowiak}, {Fuhrmann}, {Fukazawa},
  {Furniss}, {Fusco}, {Gargano}, {Gasparrini}, {Giglietto}, {Giommi},
  {Giordano}, {Giroletti}, {Glanzman}, {Godfrey}, {Grenier}, {Grove},
  {Guiriec}, {Hewitt}, {Hill}, {Horan}, {Itoh}, {J{\'o}hannesson}, {Johnson},
  {Johnson}, {Kataoka}, {Kawano}, {Krauss}, {Kuss}, {La Mura}, {Larsson},
  {Latronico}, {Leto}, {Li}, {Li}, {Longo}, {Loparco}, {Lott}, {Lovellette},
  {Lubrano}, {Madejski}, {Mayer}, {Mazziotta}, {McEnery}, {Michelson},
  {Mizuno}, {Moiseev}, {Monzani}, {Morselli}, {Moskalenko}, {Murgia}, {Nuss},
  {Ohno}, {Ohsugi}, {Ojha}, {Omodei}, {Orienti}, {Orlando}, {Paggi}, {Paneque},
  {Perkins}, {Pesce-Rollins}, {Piron}, {Pivato}, {Porter}, {Rain{\`o}},
  {Rando}, {Razzano}, {Razzaque}, {Reimer}, {Reimer}, {Romani}, {Salvetti},
  {Schaal}, {Schinzel}, {Schulz}, {Sgr{\`o}}, {Siskind}, {Sokolovsky}, {Spada},
  {Spandre}, {Spinelli}, {Stawarz}, {Suson}, {Takahashi}, {Takahashi},
  {Tanaka}, {Thayer}, {Thayer}, {Tibaldo}, {Torres}, {Torresi}, {Tosti},
  {Troja}, {Uchiyama}, {Vianello}, {Winer}, {Wood}, and
  {Zimmer}}]{2015ApJ...810...14A}
{Ackermann} M, {Ajello} M, {Atwood} WB, et~al (2015) {The Third Catalog of
  Active Galactic Nuclei Detected by the Fermi Large Area Telescope}. \apj
  810(1):14. \doi{10.1088/0004-637X/810/1/14},
  {\href{https://arxiv.org/abs/1501.06054}{{https://arxiv.org/abs/arXiv:1501.06054}}}
  {[astro-ph.HE]}

\bibitem[{{Ajello} et~al(2022){Ajello}, {Baldini}, {Ballet}, {Bastieri},
  {Becerra Gonzalez}, {Bellazzini}, {Berretta}, {Bissaldi}, {Bonino}, {Brill},
  {Bruel}, {Buson}, {Caputo}, {Caraveo}, {Cheung}, {Chiaro}, {Cibrario},
  {Ciprini}, {Crnogorcevic}, {Cutini}, {D'Ammando}, {De Gaetano}, {Di Lalla},
  {Di Venere}, {Dominguez}, {Fallah Ramazani}, {Ferrara}, {Fiori}, {Fukazawa},
  {Funk}, {Fusco}, {Gammaldi}, {Gargano}, {Garrappa}, {Gasparrini},
  {Giglietto}, {Giordano}, {Giroletti}, {Green}, {Grenier}, {Guiriec}, {Horan},
  {Hou}, {Kayanoki}, {Kuss}, {Larsson}, {Latronico}, {Lewis}, {Li}, {Liodakis},
  {Longo}, {Loparco}, {Lott}, {Lovellette}, {Lubrano}, {Madejski}, {Maldera},
  {Manfreda}, {Marti-Devesa}, {Mazziotta}, {Mereu}, {Michelson}, {Mirabal},
  {Mitthumsiri}, {Mizuno}, {Monzani}, {Morselli}, {Moskalenko}, {Negro},
  {Ojha}, {Orienti}, {Orlando}, {Ormes}, {Pei}, {Pena-Herazo}, {Persic},
  {Pesce-Rollins}, {Petrosian}, {Pillera}, {Poon}, {Porter}, {Principe},
  {Raino}, {Rando}, {Rani}, {Razzano}, {Razzaque}, {Reimer}, {Reimer},
  {Scargle}, {Scotton}, {Serini}, {Sgro}, {Siskind}, {Spandre}, {Spinelli},
  {Suson}, {Tajima}, {Torres}, {Valverde}, {Yassin}, and
  {Zaharijas}}]{2022arXiv220912070T}
{Ajello} M, {Baldini} L, {Ballet} J, et~al (2022) {The Fourth Catalog of Active
  Galactic Nuclei Detected by the Fermi Large Area Telescope: Data Release 3}.
  \apjs 263(2):24. \doi{10.3847/1538-4365/ac9523},
  {\href{https://arxiv.org/abs/2209.12070}{{https://arxiv.org/abs/arXiv:2209.12070}}}
  {[astro-ph.HE]}

\bibitem[{{Bonato} et~al(2018){Bonato}, {Liuzzo}, {Giannetti}, {Massardi}, {De
  Zotti}, {Burkutean}, {Galluzzi}, {Negrello}, {Baronchelli}, {Brand}, {Zwaan},
  {Rygl}, {Marchili}, {Klitsch}, and {Oteo}}]{2018MNRAS.478.1512B}
{Bonato} M, {Liuzzo} E, {Giannetti} A, et~al (2018) {ALMACAL IV: a catalogue of
  ALMA calibrator continuum observations}. \mnras 478(2):1512--1519.
  \doi{10.1093/mnras/sty1173},
  {\href{https://arxiv.org/abs/1805.00024}{{https://arxiv.org/abs/arXiv:1805.00024}}}
  {[astro-ph.GA]}

\bibitem[{{Buckley} et~al(2022){Buckley}, {Britto}, {Chandra}, {Krushinsky},
  {B{\"o}ttcher}, {Razzaque}, {Lipunov}, {Stalin}, {Gorbovskoy}, {Tiurina},
  {Vlasenko}, and {Kniazev}}]{2022MNRAS.517.5791B}
{Buckley} DAH, {Britto} RJ, {Chandra} S, et~al (2022) {A multiwavelength study
  of the flat-spectrum radio quasar NVSS J141922-083830 covering four flaring
  episodes}. \mnras 517(4):5791--5804. \doi{10.1093/mnras/stac2181},
  {\href{https://arxiv.org/abs/2207.14762}{{https://arxiv.org/abs/arXiv:2207.14762}}}
  {[astro-ph.HE]}

\bibitem[{{Carrasco} et~al(2015){Carrasco}, {Recillas}, {Porras},
  {Leon-Tavares}, {Chavushyan}, and {Carraminana}}]{2015ATel.7168....1C}
{Carrasco} L, {Recillas} E, {Porras} A, et~al (2015) {The ongoing Giant NIR
  flare of the Blazar candidate NVSSJ141922-083830}. The Astronomer's Telegram
  7168:1

\bibitem[{{Collioud} and {Charlot}(2009)}]{2009evga.conf...19C}
{Collioud} A, {Charlot} P (2009) {The Bordeaux VLBI Image Database}. In:
  {Bourda} G, {Charlot} P, {Collioud} A (eds) 19th European VLBI for Geodesy
  and Astrometry Working Meeting, pp 19--22

\bibitem[{{Condon} et~al(1982){Condon}, {Condon}, {Gisler}, and
  {Puschell}}]{1982ApJ...252..102C}
{Condon} JJ, {Condon} MA, {Gisler} G, et~al (1982) {Strong radio sources in
  bright spiral galaxies. II. Rapid star formation and galaxy-galaxy
  interactions.} \apj 252:102--124. \doi{10.1086/159538}

\bibitem[{{Condon} et~al(1998){Condon}, {Cotton}, {Greisen}, {Yin}, {Perley},
  {Taylor}, and {Broderick}}]{1998AJ....115.1693C}
{Condon} JJ, {Cotton} WD, {Greisen} EW, et~al (1998) {The NRAO VLA Sky Survey}.
  \aj 115(5):1693--1716. \doi{10.1086/300337}

\bibitem[{{Fermi-LAT Collaboration}(2021)}]{2021ATel15110....1F}
{Fermi-LAT Collaboration} (2021) {10-year Fermi LAT point source catalog}. The
  Astronomer's Telegram 15110:1

\bibitem[{{Gab{\'a}nyi} et~al(2018{\natexlab{a}}){Gab{\'a}nyi}, {Mo{\'o}r}, and
  {Frey}}]{2018rnls.confE..42G}
{Gab{\'a}nyi} K{\'E}, {Mo{\'o}r} A, {Frey} S (2018{\natexlab{a}}) {Infrared
  variability of radio-loud narrow-line Seyfert 1 galaxies}. In: Revisiting
  Narrow-Line Seyfert 1 Galaxies and their Place in the Universe, p~42,
  \doi{10.22323/1.328.0042}, \eprint{1807.05802}

\bibitem[{{Gab{\'a}nyi} et~al(2018{\natexlab{b}}){Gab{\'a}nyi}, {Mo{\'o}r}, and
  {Frey}}]{2018RNAAS...2..130G}
{Gab{\'a}nyi} K{\'E}, {Mo{\'o}r} A, {Frey} S (2018{\natexlab{b}}) {Mid-infrared
  Variability of the Neutrino Source Blazar TXS 0506+056}. Research Notes of
  the American Astronomical Society 2(3):130. \doi{10.3847/2515-5172/aad49f},
  {\href{https://arxiv.org/abs/1807.07462}{{https://arxiv.org/abs/arXiv:1807.07462}}}
  {[astro-ph.GA]}

\bibitem[{{Gaia Collaboration} et~al(2016){Gaia Collaboration}, {Prusti}, {de
  Bruijne}, {Brown}, {Vallenari}, {Babusiaux}, {Bailer-Jones}, {Bastian},
  {Biermann}, {Evans}, {Eyer}, {Jansen}, {Jordi}, {Klioner}, {Lammers},
  {Lindegren}, {Luri}, {Mignard}, {Milligan}, {Panem}, {Poinsignon},
  {Pourbaix}, {Randich}, {Sarri}, {Sartoretti}, {Siddiqui}, {Soubiran},
  {Valette}, {van Leeuwen}, {Walton}, {Aerts}, {Arenou}, {Cropper}, {Drimmel},
  {H{\o}g}, {Katz}, {Lattanzi}, {O'Mullane}, {Grebel}, {Holland}, {Huc},
  {Passot}, {Bramante}, {Cacciari}, {Casta{\~n}eda}, {Chaoul}, {Cheek}, {De
  Angeli}, {Fabricius}, {Guerra}, {Hern{\'a}ndez}, {Jean-Antoine-Piccolo},
  {Masana}, {Messineo}, {Mowlavi}, {Nienartowicz}, {Ord{\'o}{\~n}ez-Blanco},
  {Panuzzo}, {Portell}, {Richards}, {Riello}, {Seabroke}, {Tanga},
  {Th{\'e}venin}, {Torra}, {Els}, {Gracia-Abril}, {Comoretto},
  {Garcia-Reinaldos}, {Lock}, {Mercier}, {Altmann}, {Andrae}, {Astraatmadja},
  {Bellas-Velidis}, {Benson}, {Berthier}, {Blomme}, {Busso}, {Carry},
  {Cellino}, {Clementini}, {Cowell}, {Creevey}, {Cuypers}, {Davidson}, {De
  Ridder}, {de Torres}, {Delchambre}, {Dell'Oro}, {Ducourant}, {Fr{\'e}mat},
  {Garc{\'\i}a-Torres}, {Gosset}, {Halbwachs}, {Hambly}, {Harrison}, {Hauser},
  {Hestroffer}, {Hodgkin}, {Huckle}, {Hutton}, {Jasniewicz}, {Jordan},
  {Kontizas}, {Korn}, {Lanzafame}, {Manteiga}, {Moitinho}, {Muinonen},
  {Osinde}, {Pancino}, {Pauwels}, {Petit}, {Recio-Blanco}, {Robin}, {Sarro},
  {Siopis}, {Smith}, {Smith}, {Sozzetti}, {Thuillot}, {van Reeven}, {Viala},
  {Abbas}, {Abreu Aramburu}, {Accart}, {Aguado}, {Allan}, {Allasia},
  {Altavilla}, {{\'A}lvarez}, {Alves}, {Anderson}, {Andrei}, {Anglada Varela},
  {Antiche}, {Antoja}, {Ant{\'o}n}, {Arcay}, {Atzei}, {Ayache}, {Bach},
  {Baker}, {Balaguer-N{\'u}{\~n}ez}, {Barache}, {Barata}, {Barbier}, {Barblan},
  {Baroni}, {Barrado y Navascu{\'e}s}, {Barros}, {Barstow}, {Becciani},
  {Bellazzini}, {Bellei}, {Bello Garc{\'\i}a}, {Belokurov}, {Bendjoya},
  {Berihuete}, {Bianchi}, {Bienaym{\'e}}, {Billebaud}, {Blagorodnova},
  {Blanco-Cuaresma}, {Boch}, {Bombrun}, {Borrachero}, {Bouquillon}, {Bourda},
  {Bouy}, {Bragaglia}, {Breddels}, {Brouillet}, {Br{\"u}semeister},
  {Bucciarelli}, {Budnik}, {Burgess}, {Burgon}, {Burlacu}, {Busonero}, {Buzzi},
  {Caffau}, {Cambras}, {Campbell}, {Cancelliere}, {Cantat-Gaudin}, {Carlucci},
  {Carrasco}, {Castellani}, {Charlot}, {Charnas}, {Charvet}, {Chassat},
  {Chiavassa}, {Clotet}, {Cocozza}, {Collins}, {Collins}, {Costigan}, {Crifo},
  {Cross}, {Crosta}, {Crowley}, {Dafonte}, {Damerdji}, {Dapergolas}, {David},
  {David}, {De Cat}, {de Felice}, {de Laverny}, {De Luise}, {De March}, {de
  Martino}, {de Souza}, {Debosscher}, {del Pozo}, {Delbo}, {Delgado},
  {Delgado}, {di Marco}, {Di Matteo}, {Diakite}, {Distefano}, {Dolding}, {Dos
  Anjos}, {Drazinos}, {Dur{\'a}n}, {Dzigan}, {Ecale}, {Edvardsson}, {Enke},
  {Erdmann}, {Escolar}, {Espina}, {Evans}, {Eynard Bontemps}, {Fabre},
  {Fabrizio}, {Faigler}, {Falc{\~a}o}, {Farr{\`a}s Casas}, {Faye}, {Federici},
  {Fedorets}, {Fern{\'a}ndez-Hern{\'a}ndez}, {Fernique}, {Fienga}, {Figueras},
  {Filippi}, {Findeisen}, {Fonti}, {Fouesneau}, {Fraile}, {Fraser}, {Fuchs},
  {Furnell}, {Gai}, {Galleti}, {Galluccio}, {Garabato}, {Garc{\'\i}a-Sedano},
  {Gar{\'e}}, {Garofalo}, {Garralda}, {Gavras}, {Gerssen}, {Geyer}, {Gilmore},
  {Girona}, {Giuffrida}, {Gomes}, {Gonz{\'a}lez-Marcos},
  {Gonz{\'a}lez-N{\'u}{\~n}ez}, {Gonz{\'a}lez-Vidal}, {Granvik}, {Guerrier},
  {Guillout}, {Guiraud}, {G{\'u}rpide}, {Guti{\'e}rrez-S{\'a}nchez}, {Guy},
  {Haigron}, {Hatzidimitriou}, {Haywood}, {Heiter}, {Helmi}, {Hobbs},
  {Hofmann}, {Holl}, {Holland}, {Hunt}, {Hypki}, {Icardi}, {Irwin}, {Jevardat
  de Fombelle}, {Jofr{\'e}}, {Jonker}, {Jorissen}, {Julbe}, {Karampelas},
  {Kochoska}, {Kohley}, {Kolenberg}, {Kontizas}, {Koposov}, {Kordopatis},
  {Koubsky}, {Kowalczyk}, {Krone-Martins}, {Kudryashova}, {Kull}, {Bachchan},
  {Lacoste-Seris}, {Lanza}, {Lavigne}, {Le Poncin-Lafitte}, {Lebreton},
  {Lebzelter}, {Leccia}, {Leclerc}, {Lecoeur-Taibi}, {Lemaitre}, {Lenhardt},
  {Leroux}, {Liao}, {Licata}, {Lindstr{\o}m}, {Lister}, {Livanou}, {Lobel},
  {L{\"o}ffler}, {L{\'o}pez}, {Lopez-Lozano}, {Lorenz}, {Loureiro},
  {MacDonald}, {Magalh{\~a}es Fernandes}, {Managau}, {Mann}, {Mantelet},
  {Marchal}, {Marchant}, {Marconi}, {Marie}, {Marinoni}, {Marrese},
  {Marschalk{\'o}}, {Marshall}, {Mart{\'\i}n-Fleitas}, {Martino}, {Mary},
  {Matijevi{\v{c}}}, {Mazeh}, {McMillan}, {Messina}, {Mestre}, {Michalik},
  {Millar}, {Miranda}, {Molina}, {Molinaro}, {Molinaro}, {Moln{\'a}r},
  {Moniez}, {Montegriffo}, {Monteiro}, {Mor}, {Mora}, {Morbidelli}, {Morel},
  {Morgenthaler}, {Morley}, {Morris}, {Mulone}, {Muraveva}, {Musella},
  {Narbonne}, {Nelemans}, {Nicastro}, {Noval}, {Ord{\'e}novic},
  {Ordieres-Mer{\'e}}, {Osborne}, {Pagani}, {Pagano}, {Pailler}, {Palacin},
  {Palaversa}, {Parsons}, {Paulsen}, {Pecoraro}, {Pedrosa}, {Pentik{\"a}inen},
  {Pereira}, {Pichon}, {Piersimoni}, {Pineau}, {Plachy}, {Plum}, {Poujoulet},
  {Pr{\v{s}}a}, {Pulone}, {Ragaini}, {Rago}, {Rambaux}, {Ramos-Lerate},
  {Ranalli}, {Rauw}, {Read}, {Regibo}, {Renk}, {Reyl{\'e}}, {Ribeiro},
  {Rimoldini}, {Ripepi}, {Riva}, {Rixon}, {Roelens}, {Romero-G{\'o}mez},
  {Rowell}, {Royer}, {Rudolph}, {Ruiz-Dern}, {Sadowski}, {Sagrist{\`a}
  Sell{\'e}s}, {Sahlmann}, {Salgado}, {Salguero}, {Sarasso}, {Savietto},
  {Schnorhk}, {Schultheis}, {Sciacca}, {Segol}, {Segovia}, {Segransan},
  {Serpell}, {Shih}, {Smareglia}, {Smart}, {Smith}, {Solano}, {Solitro},
  {Sordo}, {Soria Nieto}, {Souchay}, {Spagna}, {Spoto}, {Stampa}, {Steele},
  {Steidelm{\"u}ller}, {Stephenson}, {Stoev}, {Suess}, {S{\"u}veges}, {Surdej},
  {Szabados}, {Szegedi-Elek}, {Tapiador}, {Taris}, {Tauran}, {Taylor},
  {Teixeira}, {Terrett}, {Tingley}, {Trager}, {Turon}, {Ulla}, {Utrilla},
  {Valentini}, {van Elteren}, {Van Hemelryck}, {van Leeuwen}, {Varadi},
  {Vecchiato}, {Veljanoski}, {Via}, {Vicente}, {Vogt}, {Voss}, {Votruba},
  {Voutsinas}, {Walmsley}, {Weiler}, {Weingrill}, {Werner}, {Wevers},
  {Whitehead}, {Wyrzykowski}, {Yoldas}, {{\v{Z}}erjal}, {Zucker}, {Zurbach},
  {Zwitter}, {Alecu}, {Allen}, {Allende Prieto}, {Amorim},
  {Anglada-Escud{\'e}}, {Arsenijevic}, {Azaz}, {Balm}, {Beck}, {Bernstein},
  {Bigot}, {Bijaoui}, {Blasco}, {Bonfigli}, {Bono}, {Boudreault}, {Bressan},
  {Brown}, {Brunet}, {Bunclark}, {Buonanno}, {Butkevich}, {Carret}, {Carrion},
  {Chemin}, {Ch{\'e}reau}, {Corcione}, {Darmigny}, {de Boer}, {de Teodoro}, {de
  Zeeuw}, {Delle Luche}, {Domingues}, {Dubath}, {Fodor}, {Fr{\'e}zouls},
  {Fries}, {Fustes}, {Fyfe}, {Gallardo}, {Gallegos}, {Gardiol}, {Gebran},
  {Gomboc}, {G{\'o}mez}, {Grux}, {Gueguen}, {Heyrovsky}, {Hoar}, {Iannicola},
  {Isasi Parache}, {Janotto}, {Joliet}, {Jonckheere}, {Keil}, {Kim},
  {Klagyivik}, {Klar}, {Knude}, {Kochukhov}, {Kolka}, {Kos}, {Kutka}, {Lainey},
  {LeBouquin}, {Liu}, {Loreggia}, {Makarov}, {Marseille}, {Martayan},
  {Martinez-Rubi}, {Massart}, {Meynadier}, {Mignot}, {Munari}, {Nguyen},
  {Nordlander}, {Ocvirk}, {O'Flaherty}, {Olias Sanz}, {Ortiz}, {Osorio},
  {Oszkiewicz}, {Ouzounis}, {Palmer}, {Park}, {Pasquato}, {Peltzer}, {Peralta},
  {P{\'e}turaud}, {Pieniluoma}, {Pigozzi}, {Poels}, {Prat}, {Prod'homme},
  {Raison}, {Rebordao}, {Risquez}, {Rocca-Volmerange}, {Rosen}, {Ruiz-Fuertes},
  {Russo}, {Sembay}, {Serraller Vizcaino}, {Short}, {Siebert}, {Silva},
  {Sinachopoulos}, {Slezak}, {Soffel}, {Sosnowska}, {Strai{\v{z}}ys}, {ter
  Linden}, {Terrell}, {Theil}, {Tiede}, {Troisi}, {Tsalmantza}, {Tur},
  {Vaccari}, {Vachier}, {Valles}, {Van Hamme}, {Veltz}, {Virtanen}, {Wallut},
  {Wichmann}, {Wilkinson}, {Ziaeepour}, and {Zschocke}}]{2016A&A...595A...1G}
{Gaia Collaboration}, {Prusti} T, {de Bruijne} JHJ, et~al (2016) {The Gaia
  mission}. \aap 595:A1. \doi{10.1051/0004-6361/201629272},
  {\href{https://arxiv.org/abs/1609.04153}{{https://arxiv.org/abs/arXiv:1609.04153}}}
  {[astro-ph.IM]}

\bibitem[{{Gaia Collaboration} et~al(2022){Gaia Collaboration}, {Vallenari},
  {Brown}, {Prusti}, {de Bruijne}, {Arenou}, {Babusiaux}, {Biermann},
  {Creevey}, {Ducourant}, {Evans}, {Eyer}, {Guerra}, {Hutton}, {Jordi},
  {Klioner}, {Lammers}, {Lindegren}, {Luri}, {Mignard}, {Panem}, {Pourbaix},
  {Randich}, {Sartoretti}, {Soubiran}, {Tanga}, {Walton}, {Bailer-Jones},
  {Bastian}, {Drimmel}, {Jansen}, {Katz}, {Lattanzi}, {van Leeuwen}, {Bakker},
  {Cacciari}, {Casta{\~n}eda}, {De Angeli}, {Fabricius}, {Fouesneau},
  {Fr{\'e}mat}, {Galluccio}, {Guerrier}, {Heiter}, {Masana}, {Messineo},
  {Mowlavi}, {Nicolas}, {Nienartowicz}, {Pailler}, {Panuzzo}, {Riclet}, {Roux},
  {Seabroke}, {Sordo{\o}rcit}, {Th{\'e}venin}, {Gracia-Abril}, {Portell},
  {Teyssier}, {Altmann}, {Andrae}, {Audard}, {Bellas-Velidis}, {Benson},
  {Berthier}, {Blomme}, {Burgess}, {Busonero}, {Busso}, {C{\'a}novas}, {Carry},
  {Cellino}, {Cheek}, {Clementini}, {Damerdji}, {Davidson}, {de Teodoro},
  {Nu{\~n}ez Campos}, {Delchambre}, {Dell'Oro}, {Esquej},
  {Fern{\'a}ndez-Hern{\'a}ndez}, {Fraile}, {Garabato}, {Garc{\'\i}a-Lario},
  {Gosset}, {Haigron}, {Halbwachs}, {Hambly}, {Harrison}, {Hern{\'a}ndez},
  {Hestroffer}, {Hodgkin}, {Holl}, {Jan{\ss}en}, {Jevardat de Fombelle},
  {Jordan}, {Krone-Martins}, {Lanzafame}, {L{\"o}ffler}, {Marchal}, {Marrese},
  {Moitinho}, {Muinonen}, {Osborne}, {Pancino}, {Pauwels}, {Recio-Blanco},
  {Reyl{\'e}}, {Riello}, {Rimoldini}, {Roegiers}, {Rybizki}, {Sarro}, {Siopis},
  {Smith}, {Sozzetti}, {Utrilla}, {van Leeuwen}, {Abbas}, {{\'A}brah{\'a}m},
  {Abreu Aramburu}, {Aerts}, {Aguado}, {Ajaj}, {Aldea-Montero}, {Altavilla},
  {{\'A}lvarez}, {Alves}, {Anders}, {Anderson}, {Anglada Varela}, {Antoja},
  {Baines}, {Baker}, {Balaguer-N{\'u}{\~n}ez}, {Balbinot}, {Balog}, {Barache},
  {Barbato}, {Barros}, {Barstow}, {Bartolom{\'e}}, {Bassilana}, {Bauchet},
  {Becciani}, {Bellazzini}, {Berihuete}, {Bernet}, {Bertone}, {Bianchi},
  {Binnenfeld}, {Blanco-Cuaresma}, {Blazere}, {Boch}, {Bombrun}, {Bossini},
  {Bouquillon}, {Bragaglia}, {Bramante}, {Breedt}, {Bressan}, {Brouillet},
  {Brugaletta}, {Bucciarelli}, {Burlacu}, {Butkevich}, {Buzzi}, {Caffau},
  {Cancelliere}, {Cantat-Gaudin}, {Carballo}, {Carlucci}, {Carnerero},
  {Carrasco}, {Casamiquela}, {Castellani}, {Castro-Ginard}, {Chaoul},
  {Charlot}, {Chemin}, {Chiaramida}, {Chiavassa}, {Chornay}, {Comoretto},
  {Contursi}, {Cooper}, {Cornez}, {Cowell}, {Crifo}, {Cropper}, {Crosta},
  {Crowley}, {Dafonte}, {Dapergolas}, {David}, {David}, {de Laverny}, {De
  Luise}, {De March}, {De Ridder}, {de Souza}, {de Torres}, {del Peloso}, {del
  Pozo}, {Delbo}, {Delgado}, {Delisle}, {Demouchy}, {Dharmawardena}, {Di
  Matteo}, {Diakite}, {Diener}, {Distefano}, {Dolding}, {Edvardsson}, {Enke},
  {Fabre}, {Fabrizio}, {Faigler}, {Fedorets}, {Fernique}, {Fienga}, {Figueras},
  {Fournier}, {Fouron}, {Fragkoudi}, {Gai}, {Garcia-Gutierrez},
  {Garcia-Reinaldos}, {Garc{\'\i}a-Torres}, {Garofalo}, {Gavel}, {Gavras},
  {Gerlach}, {Geyer}, {Giacobbe}, {Gilmore}, {Girona}, {Giuffrida}, {Gomel},
  {Gomez}, {Gonz{\'a}lez-N{\'u}{\~n}ez}, {Gonz{\'a}lez-Santamar{\'\i}a},
  {Gonz{\'a}lez-Vidal}, {Granvik}, {Guillout}, {Guiraud},
  {Guti{\'e}rrez-S{\'a}nchez}, {Guy}, {Hatzidimitriou}, {Hauser}, {Haywood},
  {Helmer}, {Helmi}, {Sarmiento}, {Hidalgo}, {Hilger}, {H{\l}adczuk}, {Hobbs},
  {Holland}, {Huckle}, {Jardine}, {Jasniewicz}, {Jean-Antoine Piccolo},
  {Jim{\'e}nez-Arranz}, {Jorissen}, {Juaristi Campillo}, {Julbe}, {Karbevska},
  {Kervella}, {Khanna}, {Kontizas}, {Kordopatis}, {Korn}, {K{\'o}sp{\'a}l},
  {Kostrzewa-Rutkowska}, {Kruszy{\'n}ska}, {Kun}, {Laizeau}, {Lambert},
  {Lanza}, {Lasne}, {Le Campion}, {Lebreton}, {Lebzelter}, {Leccia}, {Leclerc},
  {Lecoeur-Taibi}, {Liao}, {Licata}, {Lindstr{\o}m}, {Lister}, {Livanou},
  {Lobel}, {Lorca}, {Loup}, {Madrero Pardo}, {Magdaleno Romeo}, {Managau},
  {Mann}, {Manteiga}, {Marchant}, {Marconi}, {Marcos}, {Marcos Santos},
  {Mar{\'\i}n Pina}, {Marinoni}, {Marocco}, {Marshall}, {Polo},
  {Mart{\'\i}n-Fleitas}, {Marton}, {Mary}, {Masip}, {Massari},
  {Mastrobuono-Battisti}, {Mazeh}, {McMillan}, {Messina}, {Michalik}, {Millar},
  {Mints}, {Molina}, {Molinaro}, {Moln{\'a}r}, {Monari}, {Mongui{\'o}},
  {Montegriffo}, {Montero}, {Mor}, {Mora}, {Morbidelli}, {Morel}, {Morris},
  {Muraveva}, {Murphy}, {Musella}, {Nagy}, {Noval}, {Oca{\~n}a}, {Ogden},
  {Ordenovic}, {Osinde}, {Pagani}, {Pagano}, {Palaversa}, {Palicio},
  {Pallas-Quintela}, {Panahi}, {Payne-Wardenaar}, {Pe{\~n}alosa Esteller},
  {Penttil{\"a}}, {Pichon}, {Piersimoni}, {Pineau}, {Plachy}, {Plum}, {Poggio},
  {Pr{\v{s}}a}, {Pulone}, {Racero}, {Ragaini}, {Rainer}, {Raiteri}, {Rambaux},
  {Ramos}, {Ramos-Lerate}, {Re Fiorentin}, {Regibo}, {Richards}, {Rios Diaz},
  {Ripepi}, {Riva}, {Rix}, {Rixon}, {Robichon}, {Robin}, {Robin}, {Roelens},
  {Rogues}, {Rohrbasser}, {Romero-G{\'o}mez}, {Rowell}, {Royer}, {Ruz Mieres},
  {Rybicki}, {Sadowski}, {S{\'a}ez N{\'u}{\~n}ez}, {Sagrist{\`a} Sell{\'e}s},
  {Sahlmann}, {Salguero}, {Samaras}, {Sanchez Gimenez}, {Sanna},
  {Santove{\~n}a}, {Sarasso}, {Schultheis}, {Sciacca}, {Segol}, {Segovia},
  {S{\'e}gransan}, {Semeux}, {Shahaf}, {Siddiqui}, {Siebert}, {Siltala},
  {Silvelo}, {Slezak}, {Slezak}, {Smart}, {Snaith}, {Solano}, {Solitro},
  {Souami}, {Souchay}, {Spagna}, {Spina}, {Spoto}, {Steele},
  {Steidelm{\"u}ller}, {Stephenson}, {S{\"u}veges}, {Surdej}, {Szabados},
  {Szegedi-Elek}, {Taris}, {Taylo}, {Teixeira}, {Tolomei}, {Tonello}, {Torra},
  {Torra}, {Torralba Elipe}, {Trabucchi}, {Tsounis}, {Turon}, {Ulla}, {Unger},
  {Vaillant}, {van Dillen}, {van Reeven}, {Vanel}, {Vecchiato}, {Viala},
  {Vicente}, {Voutsinas}, {Weiler}, {Wevers}, {Wyrzykowski}, {Yoldas}, {Yvard},
  {Zhao}, {Zorec}, {Zucker}, and {Zwitter}}]{2022arXiv220800211G}
{Gaia Collaboration}, {Vallenari} A, {Brown} AGA, et~al (2022) {Gaia Data
  Release 3: Summary of the content and survey properties}. arXiv e-prints
  arXiv:2208.00211.
  {\href{https://arxiv.org/abs/2208.00211}{{https://arxiv.org/abs/arXiv:2208.00211}}}
  {[astro-ph.GA]}

\bibitem[{{Gordon} et~al(2021){Gordon}, {Boyce}, {O'Dea}, {Rudnick},
  {Andernach}, {Vantyghem}, {Baum}, {Bui}, {Dionyssiou}, {Safi-Harb}, and
  {Sander}}]{2021ApJS..255...30G}
{Gordon} YA, {Boyce} MM, {O'Dea} CP, et~al (2021) {A Quick Look at the 3 GHz
  Radio Sky. I. Source Statistics from the Very Large Array Sky Survey}. \apjs
  255(2):30. \doi{10.3847/1538-4365/ac05c0},
  {\href{https://arxiv.org/abs/2102.11753}{{https://arxiv.org/abs/arXiv:2102.11753}}}
  {[astro-ph.GA]}

\bibitem[{{H{\"o}gbom}(1974)}]{clean}
{H{\"o}gbom} JA (1974) {Aperture Synthesis with a Non-Regular Distribution of
  Interferometer Baselines}. \aaps 15:417

\bibitem[{{Hurley-Walker} et~al(2017){Hurley-Walker}, {Callingham}, {Hancock},
  {Franzen}, {Hindson}, {Kapi{\'n}ska}, {Morgan}, {Offringa}, {Wayth}, {Wu},
  {Zheng}, {Murphy}, {Bell}, {Dwarakanath}, {For}, {Gaensler},
  {Johnston-Hollitt}, {Lenc}, {Procopio}, {Staveley-Smith}, {Ekers}, {Bowman},
  {Briggs}, {Cappallo}, {Deshpande}, {Greenhill}, {Hazelton}, {Kaplan},
  {Lonsdale}, {McWhirter}, {Mitchell}, {Morales}, {Morgan}, {Oberoi}, {Ord},
  {Prabu}, {Shankar}, {Srivani}, {Subrahmanyan}, {Tingay}, {Webster},
  {Williams}, and {Williams}}]{2017MNRAS.464.1146H}
{Hurley-Walker} N, {Callingham} JR, {Hancock} PJ, et~al (2017) {GaLactic and
  Extragalactic All-sky Murchison Widefield Array (GLEAM) survey - I. A
  low-frequency extragalactic catalogue}. \mnras 464(1):1146--1167.
  \doi{10.1093/mnras/stw2337},
  {\href{https://arxiv.org/abs/1610.08318}{{https://arxiv.org/abs/arXiv:1610.08318}}}
  {[astro-ph.GA]}

\bibitem[{{Intema} et~al(2017){Intema}, {Jagannathan}, {Mooley}, and
  {Frail}}]{2017AnA...598A..78I}
{Intema} HT, {Jagannathan} P, {Mooley} KP, et~al (2017) {The GMRT 150 MHz
  all-sky radio survey. First alternative data release TGSS ADR1}. \aap
  598:A78. \doi{10.1051/0004-6361/201628536},
  {\href{https://arxiv.org/abs/1603.04368}{{https://arxiv.org/abs/arXiv:1603.04368}}}
  {[astro-ph.CO]}

\bibitem[{{Jiang} et~al(2012){Jiang}, {Zhou}, {Ho}, {Yuan}, {Wang}, {Dong},
  {Jiang}, {Ji}, and {Tian}}]{2012ApJ...759L..31J}
{Jiang} N, {Zhou} HY, {Ho} LC, et~al (2012) {Rapid Infrared Variability of
  Three Radio-loud Narrow-line Seyfert 1 Galaxies: A View from the Wide-field
  Infrared Survey Explorer}. \apjl 759(2):L31.
  \doi{10.1088/2041-8205/759/2/L31},
  {\href{https://arxiv.org/abs/1210.2800}{{https://arxiv.org/abs/arXiv:1210.2800}}}
  {[astro-ph.HE]}

\bibitem[{{Lee} et~al(2008){Lee}, {Lobanov}, {Krichbaum}, {Witzel}, {Zensus},
  {Bremer}, {Greve}, and {Grewing}}]{2008AJ....136..159L}
{Lee} SS, {Lobanov} AP, {Krichbaum} TP, et~al (2008) {A Global 86 GHz VLBI
  Survey of Compact Radio Sources}. \aj 136(1):159--180.
  \doi{10.1088/0004-6256/136/1/159},
  {\href{https://arxiv.org/abs/0803.4035}{{https://arxiv.org/abs/arXiv:0803.4035}}}
  {[astro-ph]}

\bibitem[{{Lipunov} et~al(2015){Lipunov}, {Gorbovskoy}, {Kornilov}, {Tiurina},
  {Balanutsa}, {Kuznetsov}, {Vladimirov}, {Chazov}, {Denisenko}, {Gress},
  {Ivanov}, {Budnev}, {Yazev}, {Poleshchuk}, {Konstantinov}, {Chuvalaev},
  {Tlatov}, {Dormidontov}, {Senik}, {Parkhomenko}, {Potter}, {Kniazev},
  {Kotze}, {Krushinsky}, {Zalozhnih}, {Popov}, {Bourdanov}, {Yurkov},
  {Sergienko}, {Gabovich}, {Shumkov}, and {Shurpakov}}]{2015ATel.7133....1L}
{Lipunov} V, {Gorbovskoy} E, {Kornilov} V, et~al (2015) {MASTER: very bright
  blazar flare}. The Astronomer's Telegram 7133:1

\bibitem[{{Mainzer} et~al(2014){Mainzer}, {Bauer}, {Cutri}, {Grav}, {Masiero},
  {Beck}, {Clarkson}, {Conrow}, {Dailey}, {Eisenhardt}, {Fabinsky},
  {Fajardo-Acosta}, {Fowler}, {Gelino}, {Grillmair}, {Heinrichsen}, {Kendall},
  {Kirkpatrick}, {Liu}, {Masci}, {McCallon}, {Nugent}, {Papin}, {Rice},
  {Royer}, {Ryan}, {Sevilla}, {Sonnett}, {Stevenson}, {Thompson}, {Wheelock},
  {Wiemer}, {Wittman}, {Wright}, and {Yan}}]{2014ApJ...792...30M}
{Mainzer} A, {Bauer} J, {Cutri} RM, et~al (2014) {Initial Performance of the
  NEOWISE Reactivation Mission}. \apj 792(1):30.
  \doi{10.1088/0004-637X/792/1/30},
  {\href{https://arxiv.org/abs/1406.6025}{{https://arxiv.org/abs/arXiv:1406.6025}}}
  {[astro-ph.EP]}

\bibitem[{{Meyers} et~al(2017){Meyers}, {Hurley-Walker}, {Hancock}, {Franzen},
  {Carretti}, {Staveley-Smith}, {Gaensler}, {Haverkorn}, and
  {Poppi}}]{2017PASA...34...13M}
{Meyers} BW, {Hurley-Walker} N, {Hancock} PJ, et~al (2017) {A Southern-Sky
  Total Intensity Source Catalogue at 2.3 GHz from S-Band Polarisation All-Sky
  Survey Data}. \pasa 34:e013. \doi{10.1017/pasa.2017.5},
  {\href{https://arxiv.org/abs/1701.08887}{{https://arxiv.org/abs/arXiv:1701.08887}}}
  {[astro-ph.GA]}

\bibitem[{{Murphy} et~al(2010){Murphy}, {Sadler}, {Ekers}, {Massardi},
  {Hancock}, {Mahony}, {Ricci}, {Burke-Spolaor}, {Calabretta}, {Chhetri}, {de
  Zotti}, {Edwards}, {Ekers}, {Jackson}, {Kesteven}, {Lindley}, {Newton-McGee},
  {Phillips}, {Roberts}, {Sault}, {Staveley-Smith}, {Subrahmanyan}, {Walker},
  and {Wilson}}]{2010MNRAS.402.2403M}
{Murphy} T, {Sadler} EM, {Ekers} RD, et~al (2010) {The Australia Telescope 20
  GHz Survey: the source catalogue}. \mnras 402(4):2403--2423.
  \doi{10.1111/j.1365-2966.2009.15961.x},
  {\href{https://arxiv.org/abs/0911.0002}{{https://arxiv.org/abs/arXiv:0911.0002}}}
  {[astro-ph.GA]}

\bibitem[{{Ochsenbein} et~al(2000){Ochsenbein}, {Bauer}, and
  {Marcout}}]{2000A&AS..143...23O}
{Ochsenbein} F, {Bauer} P, {Marcout} J (2000) {The VizieR database of
  astronomical catalogues}. \aaps 143:23--32. \doi{10.1051/aas:2000169},
  {\href{https://arxiv.org/abs/astro-ph/0002122}{{https://arxiv.org/abs/arXiv:astro-ph/0002122}}}
  {[astro-ph]}

\bibitem[{{Olmo-Garc{\'\i}a} et~al(2022){Olmo-Garc{\'\i}a}, {Paliya},
  {{\'A}lvarez Crespo}, {Kumar}, {Dom{\'\i}nguez}, {Gil de Paz}, and
  {S{\'a}nchez-Bl{\'a}zquez}}]{2022MNRAS.516.5702O}
{Olmo-Garc{\'\i}a} A, {Paliya} VS, {{\'A}lvarez Crespo} N, et~al (2022)
  {Optical spectroscopic characterization of Fermi blazar candidates of
  uncertain type with TNG and DOT: first results}. \mnras 516(4):5702--5711.
  \doi{10.1093/mnras/stac2640},
  {\href{https://arxiv.org/abs/2209.06518}{{https://arxiv.org/abs/arXiv:2209.06518}}}
  {[astro-ph.HE]}

\bibitem[{{Pearson}(1995)}]{modelfit}
{Pearson} TJ (1995) {Non-Imaging Data Analysis}. In: {Zensus} JA, {Diamond} PJ,
  {Napier} PJ (eds) Very Long Baseline Interferometry and the VLBA, p 267

\bibitem[{{Petrov}(2021)}]{2021AJ....161...14P}
{Petrov} L (2021) {The Wide-field VLBA Calibrator Survey: WFCS}. \aj 161(1):14.
  \doi{10.3847/1538-3881/abc4e1},
  {\href{https://arxiv.org/abs/2008.09243}{{https://arxiv.org/abs/arXiv:2008.09243}}}
  {[astro-ph.IM]}

\bibitem[{{Petrov} et~al(2011){Petrov}, {Kovalev}, {Fomalont}, and
  {Gordon}}]{2011AJ....142...35P}
{Petrov} L, {Kovalev} YY, {Fomalont} EB, et~al (2011) {The Very Long Baseline
  Array Galactic Plane Survey{\textemdash}VGaPS}. \aj 142(2):35.
  \doi{10.1088/0004-6256/142/2/35},
  {\href{https://arxiv.org/abs/1101.1460}{{https://arxiv.org/abs/arXiv:1101.1460}}}
  {[astro-ph.CO]}

\bibitem[{{Petrov} et~al(2015){Petrov}, {Natusch}, {Weston}, {McCallum},
  {Ellingsen}, and {Gulyaev}}]{2015PASP..127..516P}
{Petrov} L, {Natusch} T, {Weston} S, et~al (2015) {First Scientific VLBI
  Observations Using New Zealand 30 Meter Radio Telescope WARK30M}. \pasp
  127(952):516. \doi{10.1086/681965}

\bibitem[{{Planck Collaboration} et~al(2014){Planck Collaboration}, {Ade},
  {Aghanim}, {Armitage-Caplan}, {Arnaud}, {Ashdown}, {Atrio-Barandela},
  {Aumont}, {Aussel}, {Baccigalupi}, {Banday}, {Barreiro}, {Barrena},
  {Bartelmann}, {Bartlett}, {Battaner}, {Benabed}, {Beno{\^\i}t},
  {Benoit-L{\'e}vy}, {Bernard}, {Bersanelli}, {Bielewicz}, {Bikmaev}, {Bobin},
  {Bock}, {B{\"o}hringer}, {Bonaldi}, {Bond}, {Borrill}, {Bouchet}, {Bridges},
  {Bucher}, {Burenin}, {Burigana}, {Butler}, {Cardoso}, {Carvalho}, {Catalano},
  {Challinor}, {Chamballu}, {Chary}, {Chen}, {Chiang}, {Chiang}, {Chon},
  {Christensen}, {Churazov}, {Church}, {Clements}, {Colombi}, {Colombo},
  {Comis}, {Couchot}, {Coulais}, {Crill}, {Curto}, {Cuttaia}, {Da Silva},
  {Dahle}, {Danese}, {Davies}, {Davis}, {de Bernardis}, {de Rosa}, {de Zotti},
  {Delabrouille}, {Delouis}, {D{\'e}mocl{\`e}s}, {D{\'e}sert}, {Dickinson},
  {Diego}, {Dolag}, {Dole}, {Donzelli}, {Dor{\'e}}, {Douspis}, {Dupac},
  {Efstathiou}, {Eisenhardt}, {En{\ss}lin}, {Eriksen}, {Feroz}, {Finelli},
  {Flores-Cacho}, {Forni}, {Frailis}, {Franceschi}, {Fromenteau}, {Galeotta},
  {Ganga}, {G{\'e}nova-Santos}, {Giard}, {Giardino}, {Gilfanov},
  {Giraud-H{\'e}raud}, {Gonz{\'a}lez-Nuevo}, {G{\'o}rski}, {Grainge},
  {Gratton}, {Gregorio}, {Groeneboom}, {Gruppuso}, {Hansen}, {Hanson},
  {Harrison}, {Hempel}, {Henrot-Versill{\'e}}, {Hern{\'a}ndez-Monteagudo},
  {Herranz}, {Hildebrandt}, {Hivon}, {Hobson}, {Holmes}, {Hornstrup}, {Hovest},
  {Huffenberger}, {Hurier}, {Hurley-Walker}, {Jaffe}, {Jaffe}, {Jones},
  {Juvela}, {Keih{\"a}nen}, {Keskitalo}, {Khamitov}, {Kisner}, {Kneissl},
  {Knoche}, {Knox}, {Kunz}, {Kurki-Suonio}, {Lagache}, {L{\"a}hteenm{\"a}ki},
  {Lamarre}, {Lasenby}, {Laureijs}, {Lawrence}, {Leahy}, {Leonardi},
  {Le{\'o}n-Tavares}, {Lesgourgues}, {Li}, {Liddle}, {Liguori}, {Lilje},
  {Linden-V{\o}rnle}, {L{\'o}pez-Caniego}, {Lubin}, {Mac{\'\i}as-P{\'e}rez},
  {MacTavish}, {Maffei}, {Maino}, {Mandolesi}, {Maris}, {Marshall}, {Martin},
  {Mart{\'\i}nez-Gonz{\'a}lez}, {Masi}, {Massardi}, {Matarrese}, {Matthai},
  {Mazzotta}, {Mei}, {Meinhold}, {Melchiorri}, {Melin}, {Mendes}, {Mennella},
  {Migliaccio}, {Mikkelsen}, {Mitra}, {Miville-Desch{\^e}nes}, {Moneti},
  {Montier}, {Morgante}, {Mortlock}, {Munshi}, {Murphy}, {Naselsky}, {Nati},
  {Natoli}, {Nesvadba}, {Netterfield}, {N{\o}rgaard-Nielsen}, {Noviello},
  {Novikov}, {Novikov}, {O'Dwyer}, {Olamaie}, {Osborne}, {Oxborrow}, {Paci},
  {Pagano}, {Pajot}, {Paoletti}, {Pasian}, {Patanchon}, {Pearson}, {Perdereau},
  {Perotto}, {Perrott}, {Perrotta}, {Piacentini}, {Piat}, {Pierpaoli},
  {Pietrobon}, {Plaszczynski}, {Pointecouteau}, {Polenta}, {Ponthieu}, {Popa},
  {Poutanen}, {Pratt}, {Pr{\'e}zeau}, {Prunet}, {Puget}, {Rachen}, {Reach},
  {Rebolo}, {Reinecke}, {Remazeilles}, {Renault}, {Ricciardi}, {Riller},
  {Ristorcelli}, {Rocha}, {Rosset}, {Roudier}, {Rowan-Robinson},
  {Rubi{\~n}o-Mart{\'\i}n}, {Rumsey}, {Rusholme}, {Sandri}, {Santos},
  {Saunders}, {Savini}, {Schammel}, {Scott}, {Seiffert}, {Shellard},
  {Shimwell}, {Spencer}, {Stanford}, {Starck}, {Stolyarov}, {Stompor},
  {Sudiwala}, {Sunyaev}, {Sureau}, {Sutton}, {Suur-Uski}, {Sygnet}, {Tauber},
  {Tavagnacco}, {Terenzi}, {Toffolatti}, {Tomasi}, {Tristram}, {Tucci},
  {Tuovinen}, {T{\"u}rler}, {Umana}, {Valenziano}, {Valiviita}, {Van Tent},
  {Vibert}, {Vielva}, {Villa}, {Vittorio}, {Wade}, {Wandelt}, {White}, {White},
  {Yvon}, {Zacchei}, and {Zonca}}]{2014AnA...571A..29P}
{Planck Collaboration}, {Ade} PAR, {Aghanim} N, et~al (2014) {Planck 2013
  results. XXIX. The Planck catalogue of Sunyaev-Zeldovich sources}. \aap
  571:A29. \doi{10.1051/0004-6361/201321523},
  {\href{https://arxiv.org/abs/1303.5089}{{https://arxiv.org/abs/arXiv:1303.5089}}}
  {[astro-ph.CO]}

\bibitem[{{Readhead}(1994)}]{1994ApJ...426...51R}
{Readhead} ACS (1994) {Equipartition Brightness Temperature and the Inverse
  Compton Catastrophe}. \apj 426:51. \doi{10.1086/174038}

\bibitem[{{Sheng} et~al(2017){Sheng}, {Wang}, {Jiang}, {Yang}, {Yan}, {Dou},
  and {Peng}}]{2017ApJ...846L...7S}
{Sheng} Z, {Wang} T, {Jiang} N, et~al (2017) {Mid-infrared Variability of
  Changing-look AGNs}. \apjl 846(1):L7. \doi{10.3847/2041-8213/aa85de},
  {\href{https://arxiv.org/abs/1707.02686}{{https://arxiv.org/abs/arXiv:1707.02686}}}
  {[astro-ph.GA]}

\bibitem[{{Shepherd}(1997)}]{difmap}
{Shepherd} MC (1997) {Difmap: an Interactive Program for Synthesis Imaging}.
  In: {Hunt} G, {Payne} H (eds) Astronomical Data Analysis Software and Systems
  VI, p~77

\bibitem[{{Son} et~al(2022){Son}, {Kim}, and {Ho}}]{2022ApJ...927..107S}
{Son} S, {Kim} M, {Ho} LC (2022) {Mid-infrared Variability of Low-redshift
  Active Galactic Nuclei: Constraints on a Hot Dust Component with a Variable
  Covering Factor}. \apj 927(1):107. \doi{10.3847/1538-4357/ac4dfc},
  {\href{https://arxiv.org/abs/2201.09767}{{https://arxiv.org/abs/arXiv:2201.09767}}}
  {[astro-ph.GA]}

\bibitem[{{Wayth} et~al(2015){Wayth}, {Lenc}, {Bell}, {Callingham},
  {Dwarakanath}, {Franzen}, {For}, {Gaensler}, {Hancock}, {Hindson},
  {Hurley-Walker}, {Jackson}, {Johnston-Hollitt}, {Kapi{\'n}ska}, {McKinley},
  {Morgan}, {Offringa}, {Procopio}, {Staveley-Smith}, {Wu}, {Zheng}, {Trott},
  {Bernardi}, {Bowman}, {Briggs}, {Cappallo}, {Corey}, {Deshpande}, {Emrich},
  {Goeke}, {Greenhill}, {Hazelton}, {Kaplan}, {Kasper}, {Kratzenberg},
  {Lonsdale}, {Lynch}, {McWhirter}, {Mitchell}, {Morales}, {Morgan}, {Oberoi},
  {Ord}, {Prabu}, {Rogers}, {Roshi}, {Shankar}, {Srivani}, {Subrahmanyan},
  {Tingay}, {Waterson}, {Webster}, {Whitney}, {Williams}, and
  {Williams}}]{2015PASA...32...25W}
{Wayth} RB, {Lenc} E, {Bell} ME, et~al (2015) {GLEAM: The GaLactic and
  Extragalactic All-Sky MWA Survey}. \pasa 32:e025. \doi{10.1017/pasa.2015.26},
  {\href{https://arxiv.org/abs/1505.06041}{{https://arxiv.org/abs/arXiv:1505.06041}}}
  {[astro-ph.IM]}

\bibitem[{{Wright}(2006)}]{2006PASP..118.1711W}
{Wright} EL (2006) {A Cosmology Calculator for the World Wide Web}. \pasp
  118(850):1711--1715. \doi{10.1086/510102},
  {\href{https://arxiv.org/abs/astro-ph/0609593}{{https://arxiv.org/abs/arXiv:astro-ph/0609593}}}
  {[astro-ph]}

\bibitem[{{Wright} et~al(2010){Wright}, {Eisenhardt}, {Mainzer}, {Ressler},
  {Cutri}, {Jarrett}, {Kirkpatrick}, {Padgett}, {McMillan}, {Skrutskie},
  {Stanford}, {Cohen}, {Walker}, {Mather}, {Leisawitz}, {Gautier}, {McLean},
  {Benford}, {Lonsdale}, {Blain}, {Mendez}, {Irace}, {Duval}, {Liu}, {Royer},
  {Heinrichsen}, {Howard}, {Shannon}, {Kendall}, {Walsh}, {Larsen}, {Cardon},
  {Schick}, {Schwalm}, {Abid}, {Fabinsky}, {Naes}, and
  {Tsai}}]{2010AJ....140.1868W}
{Wright} EL, {Eisenhardt} PRM, {Mainzer} AK, et~al (2010) {The Wide-field
  Infrared Survey Explorer (WISE): Mission Description and Initial On-orbit
  Performance}. \aj 140(6):1868--1881. \doi{10.1088/0004-6256/140/6/1868},
  {\href{https://arxiv.org/abs/1008.0031}{{https://arxiv.org/abs/arXiv:1008.0031}}}
  {[astro-ph.IM]}

\bibitem[{{Yang} et~al(2018){Yang}, {Yuan}, {Yao}, {Li}, {Zhang}, {Zhou},
  {Komossa}, {Liu}, and {Jin}}]{2018MNRAS.477.5127Y}
{Yang} H, {Yuan} W, {Yao} S, et~al (2018) {SDSS J211852.96-073227.5: a new
  {\ensuremath{\gamma}}-ray flaring narrow-line Seyfert 1 galaxy}. \mnras
  477(4):5127--5138. \doi{10.1093/mnras/sty904},
  {\href{https://arxiv.org/abs/1801.03963}{{https://arxiv.org/abs/arXiv:1801.03963}}}
  {[astro-ph.HE]}

\end{thebibliography}

\begin{figure*}
  \centering
  \includegraphics[width=0.6\linewidth]{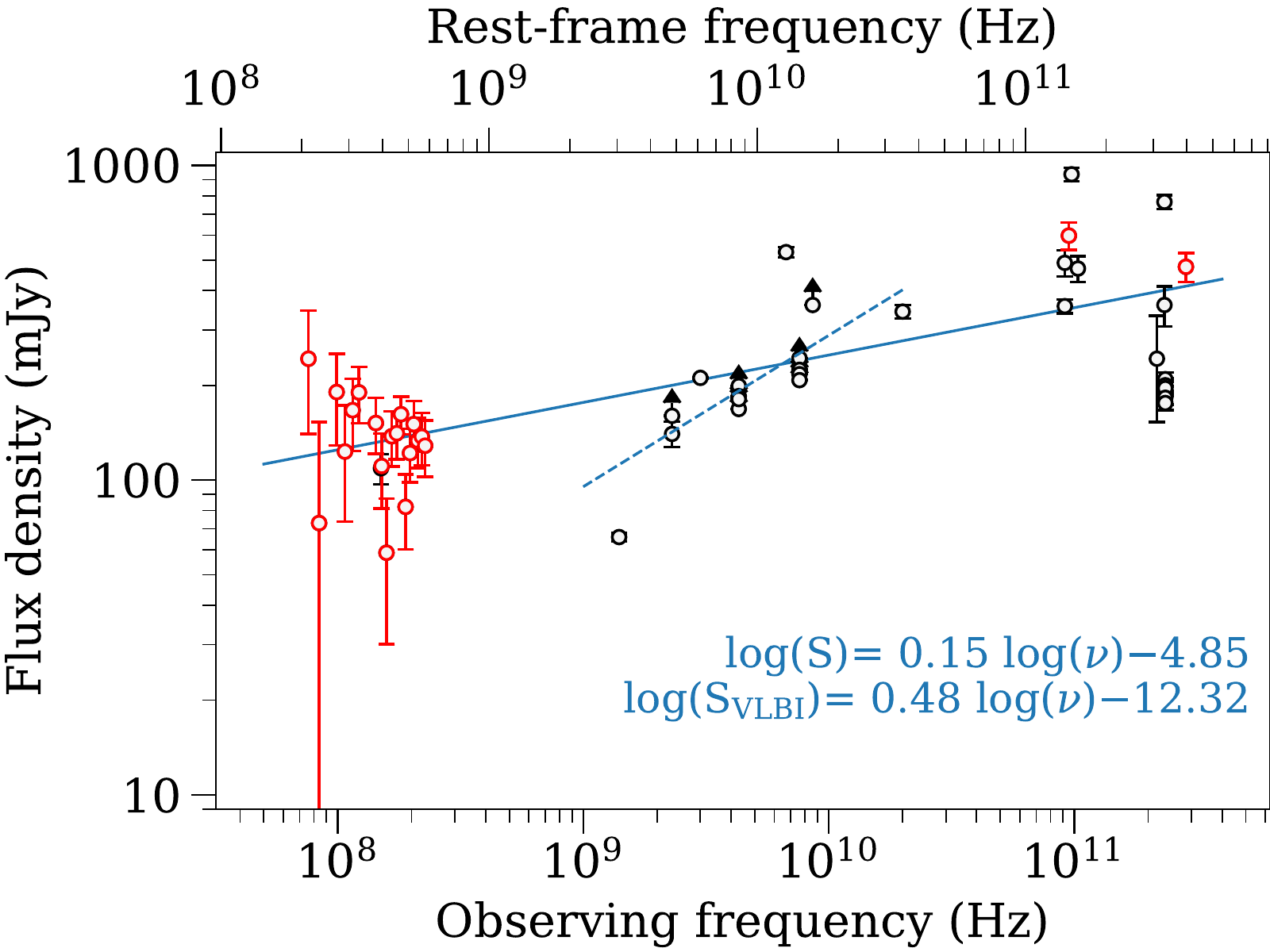}
  \caption{Radio-to-submm spectrum of J1419$-$0838. Details of the measurements are listed in Table~\ref{tab:spectrum}. Flux densities obtained from observations during the optical and $\gamma$-ray flaring events \citep{2022MNRAS.517.5791B} are plotted in red. Because of the high angular resolution, VLBI flux densities are marked as lower limits to the total flux densities. The solid and dashed blue lines denote the best-fit power-law spectra for the full data set and the VLBI data points, respectively. Note that the MWA data points (shown in red colour at low freqencies) are not included in the spectrum fit.} \label{fig:spectrum}
\end{figure*}

\begin{figure*}
  \centering
  \includegraphics[width=\linewidth]{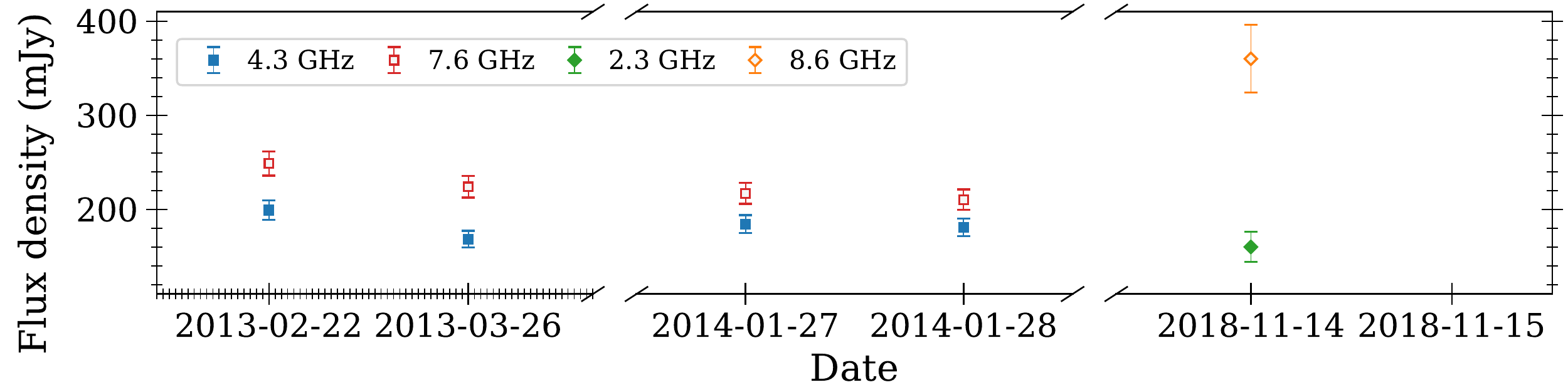}
  \caption{Radio flux density curve from VLBA observations. To avoid long gaps, the time is not displayed continuously and linearly but in distinct sections using different scaling in the horizontal axis. Minor tick marks on the leftmost side of the plot denote $1$ day, corresponding to the same time span as indicated with major tick marks in the middle and rightmost sections.  We show the flux densities from the model fitting to the 4.34- and 7.62-GHz VLBA data (Table~\ref{tab:vlbamodel}) with blue and red squares, respectively. In the absence of available visibility data and Gaussian model fits, the sum of the \textsc{clean} component flux densities at 2.3 and 8.6~GHz are taken from the Bordeaux VLBI Image Database \citep{2009evga.conf...19C}, and are shown with green and orange diamonds, respectively.}
  \label{fig:vlba_lc}
\end{figure*}

\begin{figure*}
  \centering
  \includegraphics[width=\linewidth]{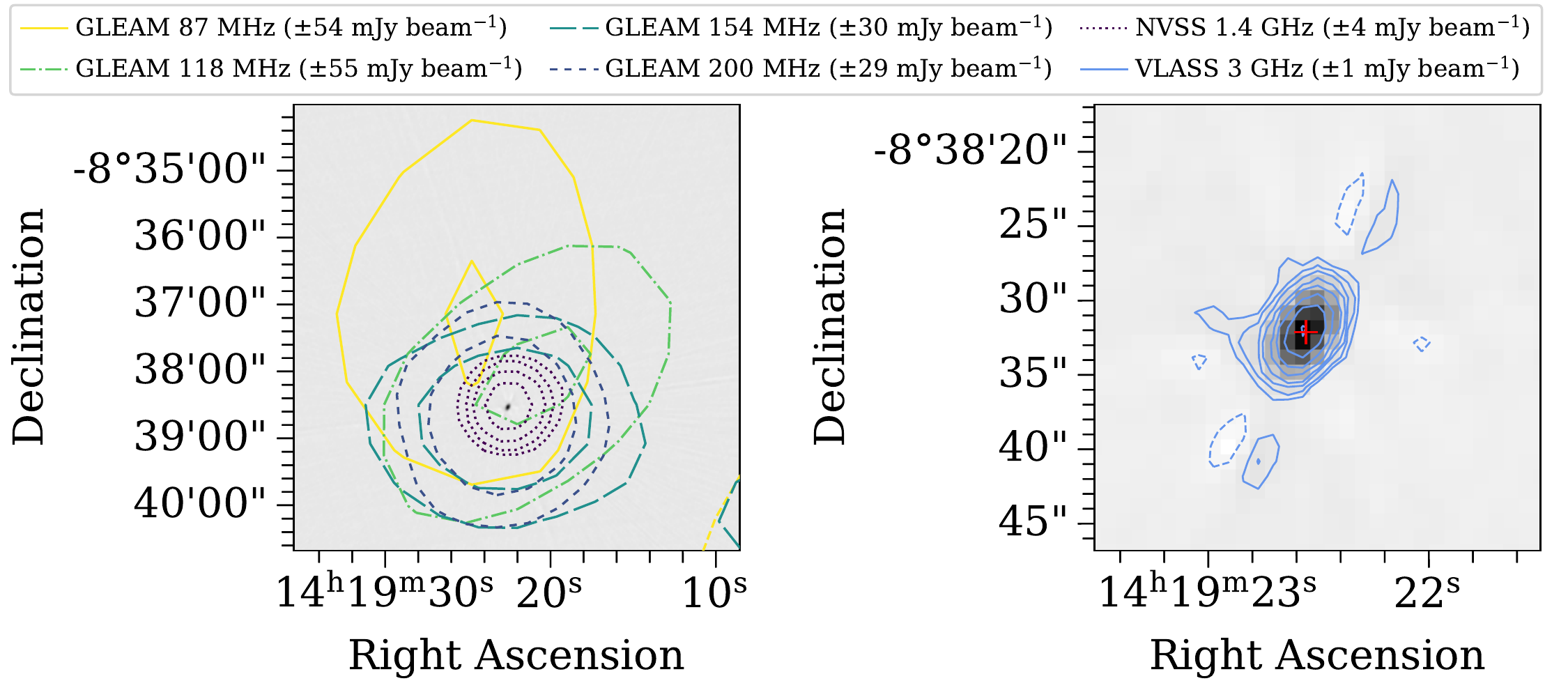}
  \caption{The large-scale radio structure of J1419--0838. In the left panel, coloured contours correspond to the GLEAM and NVSS survey brightness levels. The first contours start from $\sim3\sigma$ rms noise of the \textsc{clean} images, the values are listed in the legend. Further contour levels increase by a factor of 2. The smaller-scale VLASS radio map, shown in the right panel in more details, marks the source position in the centre. The apparent $\sim15^{\prime\prime}$ offset between the 87-MHz peak and the peaks at the other frequencies is most likely not real but can be naturally explained with the low astrometric accuracy (up to $\sim25^{\prime\prime}$) of the lowest-frequency GLEAM observations \citep{2017MNRAS.464.1146H}. In the right panel, a zoom-in to the central $30^{\prime\prime} \times 30^{\prime\prime}$ area is shown using the \textsc{clean} image from the first-epoch VLASS observations. The red cross denotes the VLBI position.}
  \label{fig:largescale}
\end{figure*}

\begin{figure*}
  \centering
 \includegraphics[width=\linewidth]{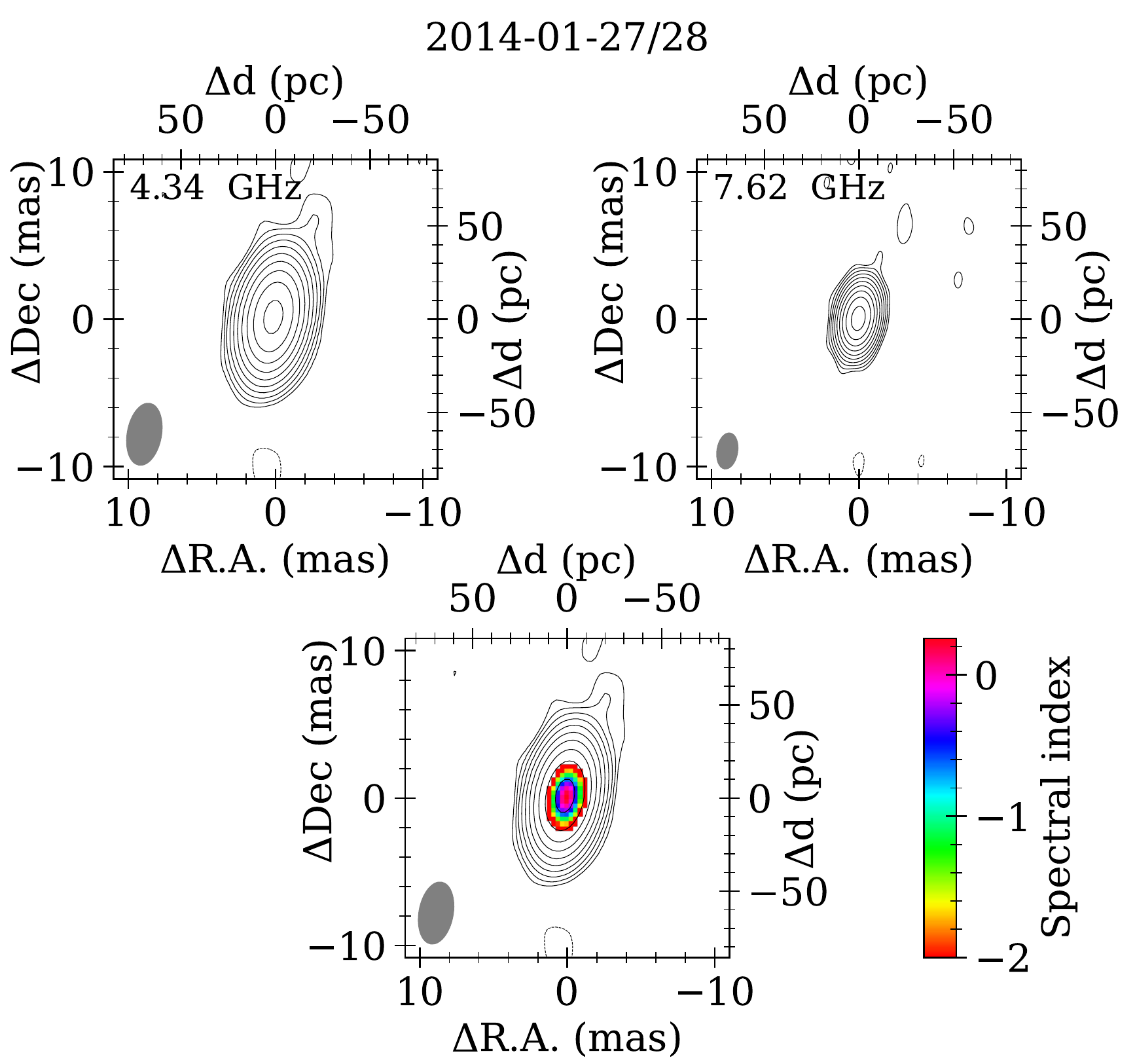}

  \caption{\textit{Upper panels:} naturally weighted \textsc{clean} images of the combined data of the 2014 VLBA observations (i.e. measurements on two consecutive days) at 4.34 and 7.62~GHz. The first contour lines are drawn at $\pm3\sigma$ rms noise, the further positive levels increase with a factor of two. The Gaussian restoring beams are shown in the bottom-left corners.  \textit{Lower panel:} spectral index map based on the combined data of the 2014 epoch VLBA observations. The colours denote the spectral index values, following the $S_\nu\propto\nu^\alpha$ convention. Contour lines denote the 4.34~GHz data. The projected linear scale is indicated in the top and right axes, the image properties are listed in Table~\ref{tab:vlbamodel}.} \label{fig:maps}
\end{figure*}

\begin{figure*}
    \centering
    \includegraphics[width=\linewidth]{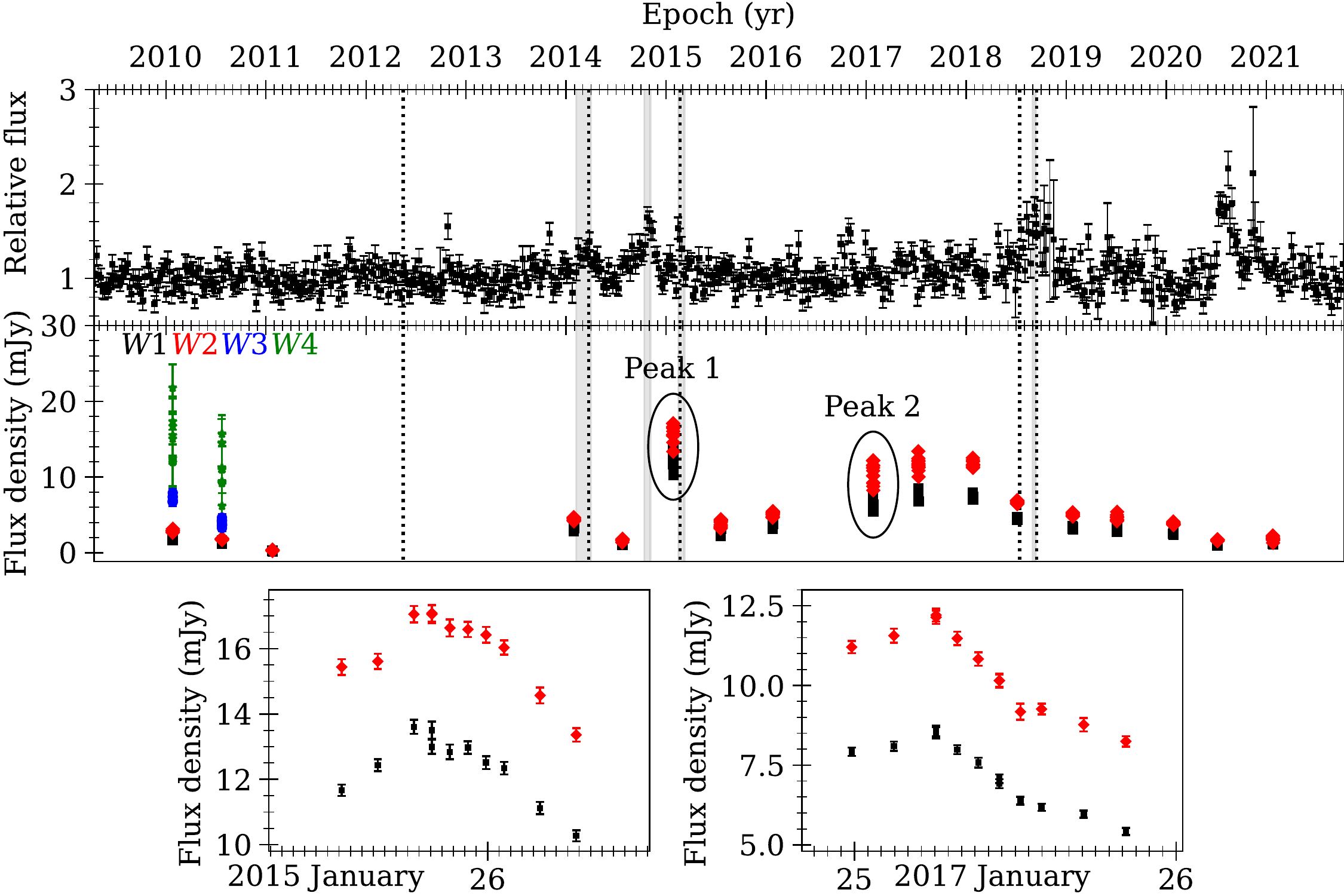}
    \caption{The $\gamma$-ray (\textit{Fermi}-LAT, 100--800 MeV; upper panel) and MIR (\textit{WISE}; middle and lower panels) light curves of the quasar J1419$-$0838 between 2010 and 2021. The grey areas and dotted vertical lines denote the flaring events reported by \citet{2022MNRAS.517.5791B} in the $\gamma$-ray and optical regimes, respectively. The two epochs of intra-day MIR-peaking events in 2015 January and 2017 January are shown in the two lower panels.}
    \label{fig:wise_lc}
\end{figure*}

\setlength{\tabcolsep}{2pt}%for table to fit on pdf page
\begin{table*}[]
\centering
\caption{Radio-to-submm spectrum}
\label{tab:spectrum}
\begin{tabular}{rrrcll}
\hline

Freq.&$S$~(mJy)&	$\sigma_S$ (mJy)	&	 Instrument 	&	 Reference 	&	Observation date(s)	\\
\hline    							
76	MHz	&	243	&	103	&	MWA	&	\citet{2017MNRAS.464.1146H}	&	2014.03.10	\\
84	MHz	&	73	&	80	&	MWA	&	\citet{2017MNRAS.464.1146H}	&	2014.03.10	\\
99	MHz	&	190	&	62	&	MWA	&	\citet{2017MNRAS.464.1146H}	&	2014.03.10	\\
107	MHz	&	123	&	49	&	MWA	&	\citet{2017MNRAS.464.1146H}	&	2014.03.10	\\
115	MHz	&	167	&	43	&	MWA	&	\citet{2017MNRAS.464.1146H}	&	2014.03.10	\\
122	MHz	&	190	&	39	&	MWA	&	\citet{2017MNRAS.464.1146H}	&	2014.03.10	\\
143	MHz	&	152	&	31	&	MWA	&	\citet{2017MNRAS.464.1146H}	&	2014.03.10	\\
150	MHz	&	109	&	12	&	GMRT	&	\citet{2017AnA...598A..78I}	&	2011.01.18.	\\
151	MHz	&	111	&	30	&	MWA	&	\citet{2017MNRAS.464.1146H}	&	2014.03.10	\\
158	MHz	&	59	&	29	&	MWA	&	\citet{2017MNRAS.464.1146H}	&	2014.03.10	\\
166	MHz	&	138	&	27	&	MWA	&	\citet{2017MNRAS.464.1146H}	&	2014.03.10	\\
174	MHz	&	141	&	25	&	MWA	&	\citet{2017MNRAS.464.1146H}	&	2014.03.10	\\
181	MHz	&	162	&	22	&	MWA	&	\citet{2017MNRAS.464.1146H}	&	2014.03.10	\\
189	MHz	&	82	&	22	&	MWA	&	\citet{2017MNRAS.464.1146H}	&	2014.03.10	\\
197	MHz	&	122	&	23	&	MWA	&	\citet{2017MNRAS.464.1146H}	&	2014.03.10	\\
204	MHz	&	151	&	27	&	MWA	&	\citet{2017MNRAS.464.1146H}	&	2014.03.10	\\
212	MHz	&	133	&	24	&	MWA	&	\citet{2017MNRAS.464.1146H}	&	2014.03.10	\\
220	MHz	&	137	&	26	&	MWA	&	\citet{2017MNRAS.464.1146H}	&	2014.03.10	\\
227	MHz	&	129	&	26	&	MWA	&	\citet{2017MNRAS.464.1146H}	&	2014.03.10	\\
1.4	GHz	&	66	&	2	&	VLA	&	\citet{1998AJ....115.1693C}	&	1993.10.23	\\
2.3	GHz	&	140	&	13	&	Parkes	&	\citet{2017PASA...34...13M}	&	2007.10–2010.01	\\
2.3	GHz	&	160	&	16	&	VLBA	&	\citet{2009evga.conf...19C}	&	2018.11.14	\\
3	GHz	&	211	&	1	&	VLASS	&	\citet{2021ApJS..255...30G}	&	2017.12.31	\\
4.3	GHz	&	199	&	10	&	VLBA	&	\citet{2021AJ....161...14P}, this work	&	2013.02.22	\\
4.3	GHz	&	168	&	9	&	VLBA	&	\citet{2021AJ....161...14P}, this work	&	2013.03.26	\\
4.3	GHz	&	185	&	10	&	VLBA	&	\citet{2021AJ....161...14P}, this work	&	2014.01.27	\\
4.3	GHz	&	184	&	12	&	VLBA	&	\citet{2021AJ....161...14P}, this work	&	2014.01.28	\\
6.7	GHz	&	530	&	20	&	Warkworth,Hobart,Ceduna&	\citet{2015PASP..127..516P}	&	2014.12.01.	\\
7.6	GHz	&	249	&	16	&	VLBA	&	\citet{2021AJ....161...14P}, this work	&	2013.02.22	\\
7.6	GHz	&	224	&	12	&	VLBA	&	\citet{2021AJ....161...14P}, this work	&	2013.03.26	\\
7.6	GHz	&	217	&	11	&	VLBA	&	\citet{2021AJ....161...14P}, this work	&	2014.01.27	\\
7.6	GHz	&	210	&	13	&	VLBA	&	\citet{2021AJ....161...14P}, this work	&	2014.01.28	\\
8.6	 GHz 	&	360	&	36	&	VLBA	&	\citet{2009evga.conf...19C}  	&	2018.11.14	\\
20	GHz	&	343	&	17	&	ATCA	&	\citet{2010MNRAS.402.2403M}	&	2007.10.26–30	\\
92	GHz	&	490	&	47	&	ALMA	&	\citet{2018MNRAS.478.1512B}	&	2015.12.26	\\
92	GHz	&	357	&	18	&	ALMA	&	\citet{2018MNRAS.478.1512B}	&	2017.12.20	\\
95	GHz	&	598	&	60	&	ALMA	&	\citet{2018MNRAS.478.1512B}	&	2015.02.20	\\
98	GHz	&	937	&	47	&	ALMA	&	\citet{2018MNRAS.478.1512B}	&	2017.08.16	\\
104	GHz	&	470	&	45	&	ALMA	&	\citet{2018MNRAS.478.1512B}	&	2015.12.26	\\
143	GHz	&	243	&	90	&	Planck	&	\citet{2014AnA...571A..29P}	&	2009.08.2–2013.10.23	\\
233	GHz	&	360	&	52	&	ALMA	&	\citet{2018MNRAS.478.1512B}	&	2015.12.26	\\
233	GHz	&	766	&	38	&	ALMA	&	\citet{2018MNRAS.478.1512B}	&	2017.07.02	\\
235	GHz	&	176	&	9	&	ALMA	&	\citet{2018MNRAS.478.1512B}	&	2016.04.10	\\
235	GHz	&	183	&	9	&	ALMA	&	\citet{2018MNRAS.478.1512B}	&	2016.04.11	\\
235	GHz	&	196	&	10	&	ALMA	&	\citet{2018MNRAS.478.1512B}	&	2016.04.13	\\
235	GHz	&	196	&	10	&	ALMA	&	\citet{2018MNRAS.478.1512B}	&	2016.04.24	\\
235	GHz	&	200	&	10	&	ALMA	&	\citet{2018MNRAS.478.1512B}	&	2016.04.24	\\
235	GHz	&	209	&	11	&	ALMA	&	\citet{2018MNRAS.478.1512B}	&	2016.05.01	\\
285	GHz	&	476	&	50	&	ALMA	&	\citet{2018MNRAS.478.1512B}	&	2015.02.20	\\

\hline
\end{tabular}\\
{\textit{Notes:} Col.~1 -- observing frequency, Col.~2 -- flux density in mJy, Col.~3 -- flux density formal error in mJy, Col.~4 -- telescope or array name; MWA -- Murchison Widefield Array, GMRT -- Giant Metrewave Radio Telescope, VLA -- Very Large Array, Parkes -- Parkes radio telescope, VLBA -- Very Long Baseline Array, WHC -- Warkworth--Hobart--Ceduna array, ALMA -- Atacama Large Millimeter/submillimeter Array, Planck -- \textit{Planck} space telescope, Col.~5 -- reference, Col.~6 -- observation date(s)} 
\end{table*}
\setlength{\tabcolsep}{6pt}%reset table column hspace

\begin{table*}[]
  \centering  
  \caption{Details of the VLBA observations}
  \label{tab:vlbaobs}
  \begin{tabular}{ccccc}
  \hline\hline
Epoch	&	Project	&	On-source time (s)	&	Participating VLBA stations	\\
	&		&		&		&		\\
\hline									
									
2013.02.22.	&	BP171a3	&	39	&	FD, HN, KP, NL, OV, SC	\\
									
2013.03.26.	&	BP171a5	&	49	&	BR, FD, HN, KP, LA, MK, NL, OV, PT, SC	\\
									
2014.01.27.	&	BP177b	&	179	&	FD, HN, LA, MK, NL, OV, PT, SC	\\
									
2014.01.28.	&	BP177c	&	160	&	FD, HN, LA, MK, NL, OV, PT, SC	\\

  \hline
  \end{tabular}
  
  {\textit{Notes:} Station codes: BR -- Brewster, FD -- Fort Davis, HN -- Hancock, KP -- Kitt Peak, LA -- Los Alamos, MK -- Mauna Kea, NL -- North Liberty, OV -- Owens Valley, PT -- Pie Town, SC -- Saint Croix}

\end{table*}

\begin{sidewaystable}
%\begin{table*}[]
  \centering
    \caption{Parameters of the models fitted to the VLBA visibility data}
  \label{tab:vlbamodel}
  \begin{tabular}{ccccccccccc}
  \hline\hline
\multirow{2}{*}{Epoch}	&	$\nu$	&	rms	&	$I_\mathrm{peak}$	&	$S$	&	$\vartheta$	&	$b_\mathrm{min}$	&	$b_\mathrm{maj}$	& $b_\mathrm{ang}$	&	$T_\mathrm{b}$	&	\multirow{2}{*}{$\delta$}\\
&	(GHz)	&	(mJy~beam$^{-1}$)	&	(mJy~beam$^{-1}$)	&	(mJy)	& (mas) &(mas) & (mas) & ($^{\circ}$) &	($10^{11}$K)	&	\\
%Epoch	&	$\nu$	&	rms	&	$I_\mathrm{peak}$	&	$S$	&	$\vartheta$	&	$b_\mathrm{min}$	&	$b_\mathrm{maj}$	& $\psi$	&	$T_\mathrm{b}$	&$\delta$\\
\hline																	
\multirow{2}{*}{2013.02.22.}																					
	&	4.34	&	0.2	&	193.2$\pm$6.2	&	199.2$\pm$10.4	&	0.43$\pm$0.01	&	1.8	&	5.2	&	10.3	&	1.3$\pm$0.2	&	2.6	\\
	&	7.62	&	0.2	&	242.2$\pm$6.7	&	249.0$\pm$15.7	&	0.23$\pm$0.01	&	1.0	&	2.8	&	11.8	&	1.9$\pm$0.2	&	3.8	\\
	&		&		&		&		&		&		&		&		&		&		\\
\multirow{2}{*}{2013.03.26.}																					
	&	4.36	&	0.2	&	166.6$\pm$5.8	&	168.2$\pm$8.8	&	$0.22\pm0.01$	&	1.7	&	4.4	&	$-5$.4	&	4.3$\pm$0.6	&	8.6	\\
	&	7.64	&	0.2	&	220.0$\pm$6.4	&	224.0$\pm$11.6	&	$0.17\pm0.01$	&	1.0	&	2.5	&	$-5.4$	&	3.2$\pm$0.3	&	6.3	\\
	&		&		&		&		&		&		&		&		&		&		\\
\multirow{2}{*}{2014.01.27.}																					
	&	4.34	&	0.1	&	182.1$\pm$4.3	&	184.8$\pm$9.5	&	$0.24\pm0.01$	&	1.7	&	4.1	&	$-1.6$	&	4.0$\pm$0.4	&	8.0	\\
	&	7.62	&	0.1	&	213.1$\pm$4.6	&	217.0$\pm$11.1	&	$0.17\pm0.01$	&	1.0	&	2.4	&	$-1.5$	&	3.0$\pm$0.3	&	5.9	\\
	&		&		&		&		&		&		&		&		&		&		\\
\multirow{2}{*}{2014.01.28.}																					
	&	4.34	&	0.1	&	180.3$\pm$4.8	&	184.2$\pm$11.7	&	$0.24\pm0.01$	&	1.7	&	4.1	&	$-3.6$	&	4.1$\pm$0.4	&	8.1	\\
	&	7.62	&	0.1	&	206.0$\pm$5.2	&	210.2$\pm$12.8	&	$0.18\pm0.01$	&	1.0	&	2.3	&	$-3.0$	&	2.5$\pm$0.3	&	5.0	\\
  \\
  
2014.01.
	&	4.34	&	0.1	&	181.6$\pm$4.2	&	183.0$\pm$11.0	&	$0.24\pm0.01$	&	2.2	&	4.2	&	$-10.2$  &	3.9$\pm$0.4	&	7.8	\\
combined	&	7.62	&	0.1	&	209.9$\pm$4.5	&	213.2$\pm$12.4	&	$0.20\pm0.01$	&	1.3	&	2.4	&	$-8.8$	&	2.1$\pm$0.2	&	4.1	\\

  \hline
  \end{tabular}
  
  {\textit{Notes:} Col.~1 -- observation epoch, Col.~2 -- observing frequency, Col.~3 -- rms noise of the image, Col.~4 -- peak intensity, Col.~5 -- flux density, Col.~6 -- size (FWHM) of the fitted circular Gaussian model component, Col.~7 -- minor axis of the elliptical Gaussian restoring beam, Col.~8 -- major axis of the restoring beam, Col.~9 -- position angle of the restoring beam major axis measured from north through east, Col.~10 -- brightness temperature, Col.~11 -- equipartition Doppler factor \citep{1994ApJ...426...51R}.}

%\end{table*}
\end{sidewaystable}
\end{document}